\documentclass[useAMS,usenatbib]{mn2e}
\usepackage{times}
\usepackage{graphicx}
\usepackage{lscape}
\usepackage{deluxetable}
\usepackage{amsmath}
\usepackage{color, colortbl}
\definecolor{Gray}{gray}{0.9}
\definecolor{LC}{rgb}{0.88,1,1}

\newcommand{\boo}{Bo\"{o}tes I}
\newcommand{\str}{Str\"{o}mgren}

\defcitealias{hug08}{HWB}
\defcitealias{gei99}{GS99}
\defcitealias{wal09}{WWJ}

\title[Photometric Metallicities]
  {Photometric Metallicities in \boo}
\author[J. Hughes et al.]
  {J. Hughes,$^1$
  G. Wallerstein,$^2$ A. Dotter,$^3$ D. Geisler$^4$ \\
  $^1$Physics Department, Seattle University, Seattle, WA 98122 \\
  $^2$Astronomy Department, University of Washington, Box 351580, Seattle,
WA 98195-1580 \\
  $^3$Research School of Astronomy \& Astrophysics, Australian National University,
  Weston, ACT 2611, Australia \\
  $^4$Grupo de Astronom\'{i}a, Departamento de Astronom\'{i}a, Universidad de Concepci\'{o}n, Casilla 160-C, Concepci\'{o}n, Chile} 
\begin{document}

\date{Accepted xxx. Received xxx.}

\maketitle
\begin{abstract}
We present new \str\ and Washington data sets for the \boo\ dwarf galaxy, and combine them with the available SDSS photometry.  The goal of this project is to
refine a ground-based, practical, accurate method to determine age and metallicity for individual stars  in \boo\ that can be selected in an unbiased imaging survey, without having to take spectra.  With few bright upper-red-giant branch stars and
distances of  about $35-250$~kpc, the ultra-faint dwarf galaxies present  observational challenges in characterizing their stellar population. Other recent studies have produced spectra and proper motions, making \boo\  an ideal test case for our photometric methods.
We produce photometric  metallicities from \str\ and Washington photometry, for stellar systems with a range of $-1.0>[Fe/H]>-3.5$.
Needing to avoid the collapse of the metallicity sensitivity of the \str\ $m_1$-index on the lower-red-giant branch, we  replace the \str\ $v$-filter with the broader Washington $C$-filter to 
minimize observing time. 
We construct two  indices: $m_*=(C-T_1)_0-(T_1-T_2)_0$, 
and 
$m_{**}=(C-b)_0-(b-y)_0$.
We find that  $CT_1by$ is the most successful filter combination, for individual stars with $[Fe/H]<-2.0$,
to maintain $\sim 0.2$~dex [Fe/H]-resolution over the whole red-giant branch. The  $m_{**}$-index would be the best choice for space-based
observations because the $(C-y)$ color is not sufficient to fix metallicity alone in an understudied system.   Our photometric metallicites of stars in the central regions of \boo\  confirm that that there is a metallicity spread of at least $-1.9>[Fe/H]>-3.7$. The best-fit Dartmouth isochrones give a mean age, for all the  \boo\  stars in our data set, of $11.5\pm 0.4$~Gyr. From ground-based telescopes, we show that the optimal filter combination is   $CT_1by$, avoiding the $v$-filter entirely. We demonstrate that we can break the isochrones' age-metallicity degeneracy with
the $CT_1by$ filters, using stars with $\log g=2.5-3.0$, which have less than a 2 per cent  change in their $(C-T_1)$-colour due to age, over a range of 10-14~Gyr.
\end{abstract}

\section{Introduction}

 The Sloan Digital Sky Survey (SDSS) survey (in ugriz bands) has been used to identify $\sim 8$ \citep[see][]{wil12} new Milky Way  satellites  \citep[for example,][]{wil05a,wil05b,bel06b,bel06a,zuc06b,zuc06a,wil10}.
 This paper is the second in a series, describing our ongoing studies of several of the recently discovered dwarf
 galaxies surrounding the Milky Way Galaxy (MWG), using the Apache Point Observatory (APO) 3.5-m telescope.  We discuss new \str\ photometry of \boo\  and  compare it with our previously-published Washington photometry and other recent spectroscopic studies, particularly those of \citet{kop11} and \citet{gil13a,gil13b}.  
 In this paper we deduce the star formation history of the central region of \boo\ from photometry, and determine the most effective and efficient combination of broad-band and medium band filters to break the age/metallicity degeneracy of populations such as these.
  
 \citet{wil10} wrote a review of the search methods, for these ``least luminous galaxies", which can be as faint as $10^{-7}$ times the luminosity of the MWG. Ten years ago, the MWG only had 11 known dwarf galaxy companions,
 which was at odds with cosmological simulations  predicting  hundreds of low mass ($10^5 M_\odot$) dark matter halos. Where were the ``missing satellites"? The apparent mismatch between the number
 of observed dark matter halos, and those predicited by the $\Lambda$CDM cosmological models was partially explained by ``simple"
 models \citep{wil10} of how stellar populations form inside low-mass dark matter halos \citep{bul00,ben02,kra04,sim07}. The first part of the problem is finding the least luminous galaxies, and the second
 issue is to determine the most efficient method to study  these sparsely-populated systems.
 A recent review by \citet{bel13} calls the pre-SDSS dwarf galaxy population ``classical" dwarfs. 

\citeauthor{wil10}'s \citeyearpar{wil10} review of the automated star-count analysis  shows how we have
 increased the completeness of unbiased sky surveys, and also describes the next generation of surveys planned for the next decade
 or so.  Detailed
 descriptions of how the automated searches were carried out can be found in \citet{wil02} and \citet*[][hereafter, WWJ]{wal09}. Once the stellar-overdensities were found, observers have to separate the dwarf spheroidal (dSph) population from that of the MWG's halo stars in the field.
 The method used by \citetalias{wal09} first selects a range of Girardi isochrones \citep[see][and references therein]{gir05}, assuming that the dSphs have populations which are aged between
 8 and 14~Gyr, with $-1.5<[Fe/H]<-2.3$. This range of models was used to create a colour-magnitude (CM) filter, which was then moved to
 16 values of the distance modulus, between 16.5 and 24.0. The software then looked for stellar overdensities, above a certain
 detection threshold;  \citetalias{wal09} describe this in detail, along with how the data was simulated. Along with dwarf galaxies, the MWG's halo has tidal debris and
  unbound star clusters, which can also be picked up in this method. Following up the detections with photometry
 and spectroscopy is essential to finding which of the detections are actual dwarf galaxies.
 When these SDSS searches were performed, \citet{wil10} notes that the least luminous galaxies can only be detected
 out to about 50 kpc. The CM filter method \citepalias{wal09} can locate systems with distances in the
 range of 20-600 kpc, but it is brightness limited.  \citet{kop08} and \citet{bel13} discuss the SDSS completeness limits, where dwarf satellite of our Galaxy are complete out to a virial radius of 280 kpc at $M_V\sim -5$ (using SDSS DR5). For systems such as Segue 1 with $M_V\sim -3$, only a few percent of the ``virial volume" can be sampled. Thus, part of the problem is the faintness of the ``darker" satellites, and part of it is the automated detection method uses filters which do not separate the sparse dwarf populations from the foreground stars in colour-magnitude diagrams.

Some of these lowest luminosity dSphs (discovered in the SDSS) do not look like the tidal debris of collisions \citep[as discussed in][]{bel13}, they look like the primordial leftovers from galaxy-assembly and \citet{gil13a} go as far as to identify \boo\ as ``surviving example of one of the first bound objects to form in the Universe." 

Early star formation can progress in different ways, depending on the star formation rate, but the pathway should be detectable in spectroscopic surveys \citep{gil13a,gil13b}. Both papers discuss two star formation channels for these extremely metal-poor systems. Rapid-star-formation after Pop III core-collapse supernovae  can produce carbon-rich (CEMP-no) stars. The ``long-lived, low star-formation rate"  would produce more carbon-normal abundances. \citet{gil13a,gil13b} identify \boo\ as the latter case. However, some authors have identified a few red giant stars in Boo I as carbon-rich \citep{lai11,gil13b}. The $2.5\times r_h$ distance of the radial-velocity-confirmed member, Boo-1137, from the center of \boo\  might indicate that a much more massive original system is being stripped (see Figure 1). The half-light radius of \boo\ is about 240 pc \citep{gil13a}.  At first examination, \boo\ appears to be a normal, if extended, dSph at a Galactocentric distance of about 60 kpc, with $e\sim 0.2$, and $-3.7$ to (at least) $-1.9$ in [Fe/H].  Any stellar system/dwarf galaxy found at around 20 kpc would be contaminated by MWG thick disk and halo stars, while those which lie beyond 100 kpc are mostly affected by the MWG halo stars. However, the Sloan filters  do not separate out
 the dSph stars from the MWG stars very well on colour-magnitude diagrams (CMDs) or colour-colour plots \citep[][hereafter, HWB]{bel06a,hug08}.

\begin{figure} 
\includegraphics[scale=1.0,width=84mm]{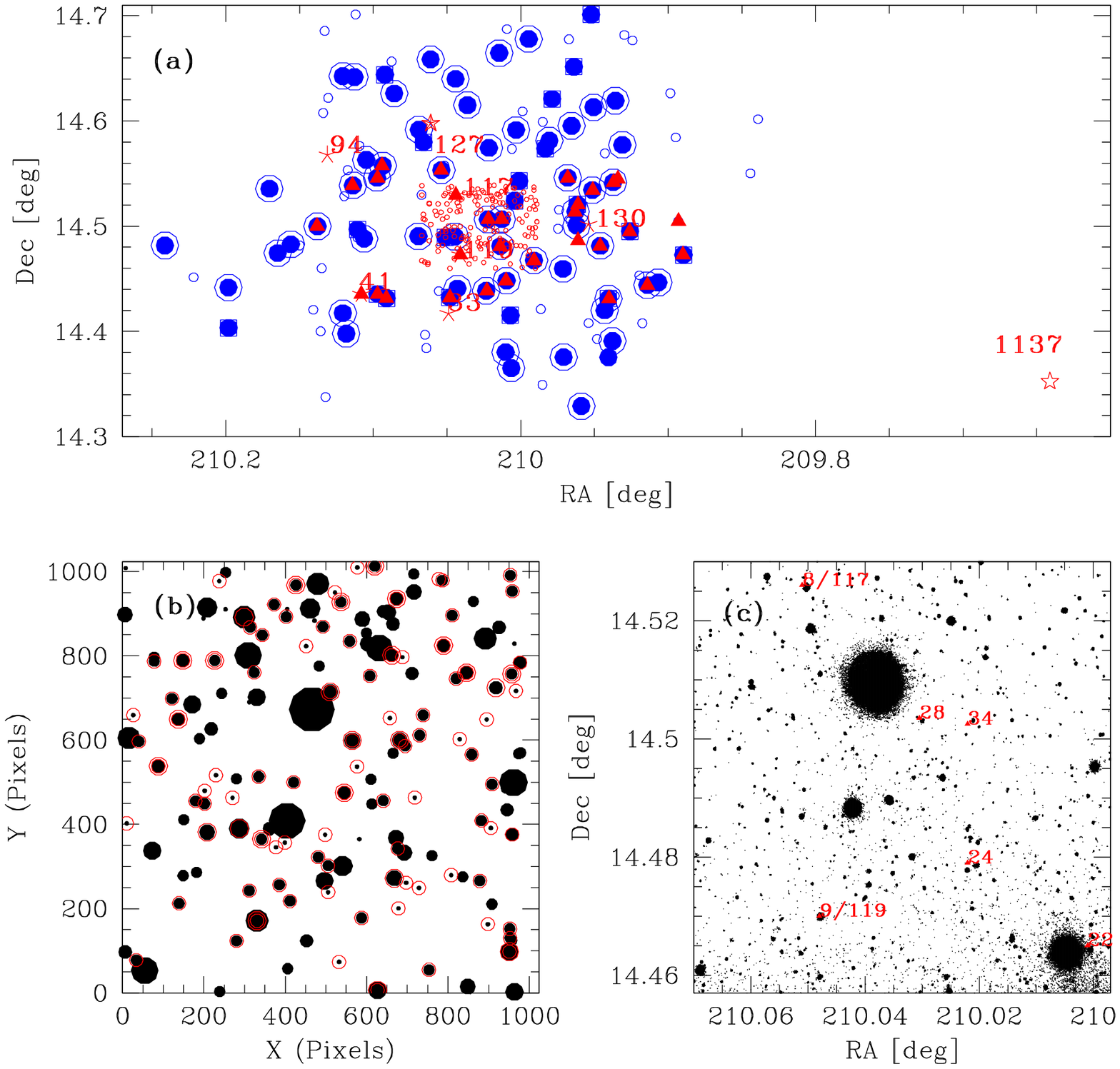}
\caption{
\bf (a)  \rm Map of the stars on the \boo\ region. Stars with radial velocity measurements and identified as  members, from \citet{mar07} are shown as filled red triangles. \citeauthor{kop11}'s \citeyearpar{kop11} sample is shown as open blue circles, and the stars having radial velocities consistent with the dSph are indicated as filled blue circles (74). In addition, the stars with velocities $95<V_r<108$ km/s are encircled by an outer blue ring (55), and those with a greater dispersion (19), but within the range $85<V_r<119$ km/s are the filled blue squares. The stars with high resolution spectroscopy discussed by \citet{nor09,nor10a,nor10b} and \citet{fel09} are numbered as Boo-1137 and Boo-127 as open red stars. The 7 RGB stars studied by \citet{gil13b} are indicated by red 5-point-line stars. The small red open circles are the 165 stars from \citetalias{hug08}, and show the $4.78 \times 4.78$ square-arcminute, APO SPIcam FOV. We also re-observed the outlying RGB stars in $vbyCT_1T_2(RI)$.
\bf (b) \rm Finding chart for \citetalias{hug08}'s data on a $1024\times 1024$ pixel scale. The black filled circles are proportional to $T_1$-magnitudes, and the open circles around them signify stars statistically likely to be Boo I members from color-comparisons. 
\bf (c) \rm The central field co-added image of the same field as (b), with the RGB stars from \citetalias{hug08} and this paper identified.}
\end{figure}

 \citet{kop11} used an ``enhanced" data reduction technique to achieve velocity errors of better than 1 km/s with the fiber-fed VLT/FLAMES+GIRAFFE system, around the
$Ca{II}$-triplet (hereafter CaT, 8498, 8542, and 8662\AA ).
  \citeauthor{kop11}'s \citeyearpar{kop11} interpretation of their data prefers a two-component velocity distribution for \boo , with 98\% confidence. We plot the finding chart for this survey, our photometric data and \citeauthor{mar07}'s \citeyearpar{mar07} survey in Figure 1. \citet{kop11} state that it is less likely that there is a one-component Gaussian with a velocity dispersion of $4.6^{+0.8}_{-0.6}$~km/s. About 70\% of the stars in their data set 
have a velocity dispersion of $2.4^{+0.9}_{-0.7}$~km/s (the ``colder" population) and a ``hotter" population with a velocity dispersion of $~9$~km/s. They give an alternative explanation that \boo\ could have a one component velocity distribution, but that the stars' velocities
are not distributed isotropically; we agree with \citet{kop11} that this model is hard to test without full spatial coverage.
From Figure 1, it is clear that a much deeper survey needs to be made of the whole region out to at least 3 half-light radii, as \citet{kop11} assert, but that there is likely to be a very low density of  ``halo'' objects belonging to \boo . The multiple short exposures around CaT, used for the \citet{kop11} study, can't resolve metallicities below $[Fe/H]\sim -2.5$. High-resolution spectroscopy of these stars require about 15 hours observation each with VLT FLAMES and GIRAFFE and FLAMES \citep{gil13a,gil13b}.

If we have any hope of mapping the full extent of \boo\ and examining the more distant systems which are likely to be found in the future, we need a more efficient method to identify age and metallicity spreads in sparsely populated systems, before selecting stars for spectroscopy. \citet{mar07} found only 30/96 stars identified as having the appropriate SDSS colors had \boo 's radial velocity. SDSS $ugriz$-filters were not designed for this task. In this paper, we are using the relatively well-studied \boo\ to find an efficient photometric method of locating dSph-members and solving for age and metallicity, with at least the $0.5~dex$ accuracy in $[Fe/H]$ given by the CaT spectra. 
Simply stated, our problem in studying the stellar populations in the dSphs/ultra-faint dwarfs (UFDs), is that some have few or no upper-red-giant branch stars. Without these bright stars, we require exposure-times of many thousands of seconds to achieve acceptable S/N, when observing in 
 blue or UV filters. If we want to survey these objects spectroscopically, we should have an efficient way of identifying interesting stars by color, over and above the SDSS photometry.
 Traditional gravity-sensitive and metallicity-sensitive colours and indices involve the use of filters which become impractical with red, faint stars  on the subgiant branch (SGB). Which blue filter is best for balancing
 metallicity sensitivity with achievable S/N? Our method for comparing spectroscopic and photometric metallicity measurements is set out in \S 2, with the observations described in \S 3. The detailed analysis is given in \S 4 and \S 5.

\section{Method}

\it How do we characterize the stellar populations of a system with similar properties to  \boo ? \rm Studies show \citep[][and references therein]{wil10} that the majority of these dSph and UFDs have very few  red-giant branch (RGB) stars, which are normally the only stars bright enough for high-resolution spectroscopy.

\subsection{Filters}

We considered using some combination of the Washington, \str , and SDSS filters, which are available at most observatories (see Figure 2).
We began this project in 2007,  imaging several of the dwarf galaxies using the Washington $CT_1T_2$-filters (using R \& I instead of $T_1$ and $T_2$ to reduce observing time;
see \S2.3) in 2007, and the first paper on \boo\ has been published
 \citepalias{hug08}. The second, and concurrent, part of the imaging project began 
 in early 2008, utilizing the \str\ $vby$-filters. All the observations used for the analysis in this
 paper are given in Table~1, and discussed in detail in \S 3.
 
 Our method used an evolving choice of filters, and the
 early part of the \str\ work was described in \citet{hug11a,hug11b}. Strong evidence of a spread in $[Fe/H]$ came from early spectroscopy \citep{mar07},  from our Washington observations \citepalias{hug08}, and higher resolution spectroscopy by \citet{nor08,nor10a,nor10b,lai11,gil13b}. 

The Washington system was used to define the \citet[][hereafter, GS99]{gei99} standard giant branches. \citetalias{gei99} show that compared to \citet{arm90}, using
the broad-band $V$- \& $I$-filters, they can obtain three times the precision in metallicity determinations,
at about a magnitude below the tip of the RGB, at around $M_{T_1}=-2.$ 

\subsection{Metallicity Scales}

\citet{mar07} and \citetalias{hug08}  found evidence of metallicity spread in  \boo , which has been confirmed by higher resolution spectroscopy \citep{nor08,iva13,gil13a,gil13b}. \citetalias{hug08}'s estimate of the spread in $[Fe/H]$ for \boo\ was calibrated to \citetalias{gei99}'s standard giant branches, and is therefore tied to the metallicity-scale of globular clusters used in that paper.  \citet{sie06} notes that Boo I's stellar population is similar to that of M92
\citetalias[as used in][]{hug08}, and we note that  M92 and M15 are regarded as the most metal poor 
globular clusters at $[Fe/H]\sim -2.3$.  \citetalias{gei99} discuss the metallicity scales of \citet{zin85,zin84,car97},  and also define a ``HDS" scale of their own, which takes the unweighted means of available high-dispersion spectroscopy (mostly from \citeauthor{rut97}'s \citeyearpar{rut97} study of calcium-triplet strengths). In \citetalias{gei99}, the most metal-poor globular cluster in their study, M15,  has  $[Fe/H]=-2.24$ on the HDS scale, $[Fe/H]=-2.15$ on the \citet{zin84} scale, but
-2.02 on the \citet{car97} calibration. Within the uncertainties, this magnitude of disagreement alone could explain the difference
in mean [Fe/H] between the Washington photometry and the SDSS data \citep[also see discussion in][]{hug11a}. The Washington filters
and the \citetalias{gei99} standard giant branches are generally designed to return the CaT-matched metallicity scale of \citet{zin85}. \citetalias{gei99} derive nine calibrations based on 
 $M_{T_1}$  and metallicity, and let the user decide which is appropriate for their cluster/galaxy.

 The \citetalias{hug08}-average value of $[Fe/H]=-2.1^{+0.3}_{-0.5}$~dex\footnote{which we normally quote
 as $\pm 0.4$ dex, but the error bars combined with the calibrations make the metallicity determination slightly asymmetric.} was determined from the 7 brightest members of \boo\ in their data set (detected in the $CRI$ filters), with the smallest photometric uncertainties. \citet{hug11a} discuss a recent paper by \citet{lai11} on \boo , which 
used low resolution spectra,  the SDSS bands and other available filters to characterize the stars. \citet{lai11} determined [Fe/H], [C/Fe], and [$\alpha$/Fe] for each target star, utilizing a 
new version of the SEGUE Stellar Parameter Pipeline \citep[SSPP;][]{lee08a,lee08b} named the n-SSPP (the method for non-SEGUE data).  The \citet{lai11} study found the [Fe/H]-range to be  about $2.0-2.5$ dex, and a mean $[Fe/H]=-2.59$ (with an uncertainty of $0.2$ dex in each measurement). \citetalias{hug08}, using Washington photometry alone, find $[Fe/H]=-2.1$, and a range $> 1.0$ dex in the central regions. \citet{mar07} studied 30 objects in \boo\ and found the same mean value as \citetalias{hug08} with the calcium triplet (CaT) method. It is known that 
the CaT-calibration may skew to higher [Fe/H]-values at the lower-metallicity end, below $[Fe/H]\sim-2.0$ \citep{kir08}. \citet{kop11} comment that the inner regions of \boo\ do seem to be more metal-rich at the $2.4\sigma$ level, than the outer regions (Figure 1a), which our photometry does not cover. \citet{kop11} examined 16 stars from \citet{nor10a} and showed that progressing radially outwards from the center of Figure 1a, the inner 8 stars have a mean $[Fe/H] =  -2.30 \pm 0.12$ and the outer 8 have $[Fe/H] = -2.78 \pm 0.17$.

 \citet{fre11},  have amassed  a  high-resolution spectroscopic study of the chemical composition of several UFDs, and a recent paper by \citet{kir12} discusses how supernovae (SN) enrich/pollute the gas in low-mass dSphs. In the latter paper, they comment that SN in systems like \boo\ would be more effective at enrichment,
on an individual basis, than early massive stars were at enriching the MW's halo because there
was less gas to contaminate. In addition, \citet{kir12} note that a star in a dSph with $[Fe/H]\sim -3.0$ is  sampling the previous generation of massive stars with $[Fe/H]<<-3.0$. This is a
particularly important point when we consider that \citet{nor10b} have found that Boo-1137 has $[Fe/H]=-3.7$, and this is discussed at length in \citet{gil13a,gil13b}.

\subsection{Practical Filter Sets for Studying Nearby Dwarf Galaxies}

\citeauthor{hug11a} \citeyearpar[][a summary of a conference presentation]{hug11a} discussed recent papers that explored the optimal colour-pairs to use for age and metallicity studies \citep[e.g.][]{li08,hol11}. However, much of the rhetoric is theoretical
and involves testing on nearby, densely populated globular clusters. The search for \it practical \rm colour-pairs also challenges the observer to use filters that can be employed on the same instrument, on the
same night (if possible), to minimize zero-point offsets and seeing differences.

\citet{ros14} calibrated the Dartmouth isochrones for Hubble Space Telescope (HST)/Wide Field Camera 3 (WFC3) using 5 globular clusters in the metallicity range $-2.30<[Fe/H]<+0.4$. They found that clusters with known distances, reddening and ages could have their metallicities determined to $\sim 1.0$~dex (overall). Otherwise, non-pre-judged results on the globulars' dominant metallicity showed  the best colors to be: $F336W-F555W$ (SDSS-u combined with Johnson-V) yields the cluster metallicity to $\sim 0.2$ to $0.5$~dex (high to low metallicity), $F390M-F555W$ ($Ca_{II}$ Cont. combined with Johnson-V) gives $\sim 0.15$ to $0.25$~dex, and $F390W-F555W$ (Washington-C and Johnson-V) gives
$\sim 0.2$ to $0.4$~dex. In this paper, we did not test $F390M$, but $(C-y)$ is equivalent to $(C-V)$. With the dSphs, the systems are not very well-studied, and we require the best color for individual stars, not the whole RGB.

\citet{cal12} have produced a metallicity calibration for dwarf stars based on the \str\ $m_1$-index and near-infrared colours, but their calibration works better for populations which are more metal-rich than the UFDs.
In this section, we compare the various filter-combinations and note some pitfalls which may be unique to UFD populations.
Figure 2a shows the transmission curves for the filters given in Table~2 \citep{bes05}, taken from the CTIO website\footnote{http://www.ctio.noao.edu/instruments/filters/index.html}, with the ATLAS9\footnote{http://wwwuser.oat.ts.astro.it/castelli/grids.html and \citet{cas03,cas04}.} model flux density for a star with $T_{eff}=4750 K$, $[Fe/H]=-2.5$, $[\alpha/Fe]=+0.4$, $\log g=1.5$.  
\begin{figure} 
\includegraphics[scale=1.0,width=84mm]{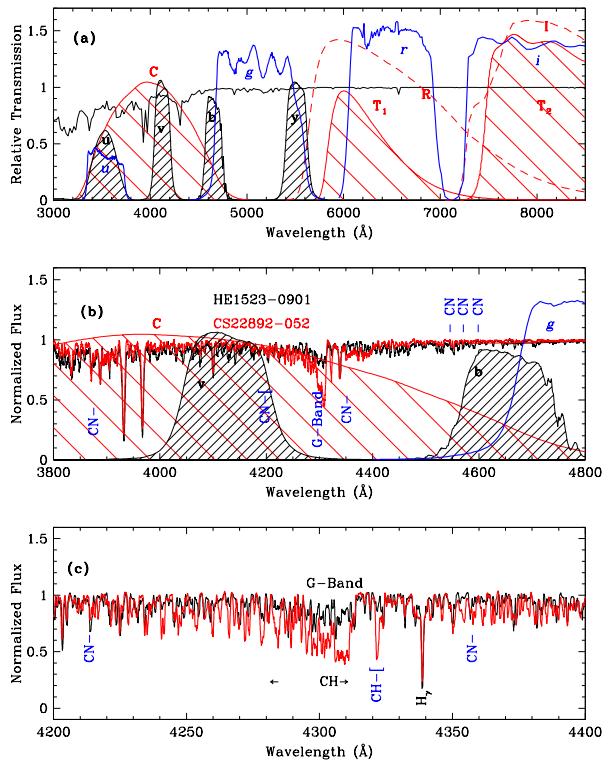}
\caption{
 \bf (a) \rm Transmission curves for the filters given in Table~2, from the CTIO website. We also show
the ATLAS9 model flux density for $T_{eff}=4750 K$, $[Fe/H]=-2.5$, $[\alpha/Fe]=+0.4$, $\log g=1.5$. Str\"{o}mgren filters (including $u$) are shown as shaded black curves. Washington filters are shown in shaded red, with the $R-$ and $I$-filters as dashed red lines. SDSS filters are shown in blue.
\bf (b) \rm Normalized flux plots for HE1523-0901 (black) and CS22892-052 (red), from 3800-4800\AA . We show the filter
transmission curves for $C$ (red), $v$ (black), $b$ (black), and $g$ (blue) filters; we  note the major CN and CH features. The original resolution has been smoothed to show the carbon-sensitive absorption.
\bf (c) \rm The normalized flux curves for HE1523-0901 (black) and CS22892-052 (red), from 4200-4400\AA , with major CN \& CH spectral features marked. }
\end{figure}

Overall, the best photometric system designed for separating stars by metallicity is considered to be the intermediate-band Str\"{o}mgren photometry \citep{str66}. The distant RGB stars in the dSphs are very faint at Str\"{o}mgren-u, and only the 8 brightest proper-motion members are detected at SDSS-u (6 from \citetalias{hug08} and the 2 extra RGB stars with high-resolution spectra, including Boo-1137). 
Without the $u$-band, we are unable to obtain the surface gravity-sensitive, $c_1$-index, where \begin{equation}
c_1=(u-v)-(v-b),
\end{equation} 
which measures the Balmer jump.

The metallicity of the stars ($Fe$ plus light elements) is sensitive to the $m_1$-index, where 
\begin{equation}
m_1=(v-b)-(b-y).
\end{equation} 

The $(b-y)$-colour  is a measure of the temperature and $(v-b)$ is a measure of metallic line blanketing (see Figure 2a). Many papers have mapped the Str\"{o}mgren metallicity index to [Fe/H] \citep[e.g][]{hil00,cal07,cal09} and find that calibrations fail for the RGB stars at $(b - y) < 0.5$ for all schemes. \citet{far07} commented that the loci of metal-rich and metal-poor stars overlap  on the lower-RGB, which they say is likely due to the larger photometric errors. Although 
this statement is not false, it is not the only reason for the issue  (Figure 2). The $m_1$-index loses sensitivity as the difference in line absorption between $b$ and $v$ becomes equal to the difference in line absorption between $b$ and $y$ \citep{hug11a}. As stars become fainter lower down the RGB, the surface temperature rises and the lines get weaker \citep[also see:][and references therin]{one09,arn10}. The latter paper discusses the \citet{van06} isochrones and the temperature-colour transformation by \citet{cle04}, and makes a point that
their classification scheme can only be used for giants with $(b - y)_0 \geq 0.6$.

Also from Figure 2, we can see  the advantages that the Washington filters provide over the \str\ and SDSS filters. The broad $C$-filter includes the metallicity-defining lines contained in the narrower $v$-filter and part of the $b$-filter, and also 
surface-gravity sensitive Str\"{o}mgren-$u$ and SDSS-$u$. Thus, the colour $(C-T_1)$ should be sensitive to $T_{eff}$, $[Fe/H]$, $[\alpha/Fe]$, and $\log g$ \citepalias{gei99}. The Str\"{o}mgren filters are more  effective than Washington bands in a system with a well-populated upper RGB, or if the stellar system is close enough to have $\sim 1$ per cent  photometry below the sub-giant branch (SGB), where the isochrones separate.
As discussed in \citetalias{hug08}, \citet{gei96} and \citet{gei91}, the more-commonly used broadband $R-$ and $I$-filters can be converted linearly to Washington $T_1$ and $T_2$, but with less observing time needed (also see the filter profiles in Figure 2a). The $C$-filter is broader than the Johnson $B$-band, and is more sensitive to line-blanketing. Washington-$C$ is a better filter choice than Johnson-$B$ or  Str\"{o}mgren-$v$ for determining metallicity in faint, distant galaxies. Table 2 includes estimates for
the total exposure times required to reach the main-sequence turn-off (MSTO) of dSphs with the WFPC3 on HST (also see \citealp{ros14}).

Summarizing  comments by \citet{sne03}, metallicity is usually synonymous with [Fe/H], but
other elements may be inhomogeneously-variable in dSphs as well as the Milky Way's halo.
\begin{equation}
[Fe/H]=\log_{10} ({N_{Fe}/ N_H})_* - \log_{10} ({N_{Fe}/ N_H})_\odot .
\end{equation}
The \it metallicity \rm is normally taken to be:
\begin{equation}
Z=Z_0 (0.694f_\alpha + 0.306),
\end{equation}
where $f_\alpha \equiv [\alpha /Fe]$, the $\alpha$-enhancement factor, and $Z_0$ is the ``heavy element abundance by mass for the solar mixture with the same $[Fe/H]$" \citep{kim02}.

In Figure 2b and 2c, we use 2 metal-poor RGB stars to illustrate the sensitivity of the \str , Washington and SDSS filters to carbon-enhancement. HE 1523-0901 \citep[black line:][]{fre07} is a r-process-enhanced metal-poor star 
with $[Fe/H]\approx -3.0$, $[C/Fe]=-0.3$, $\log T_{eff}=4650K$, and $log \;g=1.0$. CS 22892-052 (red line) is also an r-process rich object \citep{sne03,sne09,cow11} with $[Fe/H]\approx -3.0$, $[C/Fe]\approx 1.0$, $\log T_{eff}=4800K$, $log \;g=1.5$, and
$[\alpha /Fe]\approx +0.3$. The change in the CH-caused G-band is apparent in Figure 2c, and CN/CH features affect the $C$ (in particular), $v-$, and $b$-filters, but the SDSS $g$-band is relatively clear of contamination, but $g$ is not  very sensitive to metallicity either. The spectra shown in Figure 2 were provided by Anna Frebel (private communication).

 \citet{car11} reports a study of globular cluster stars with CN/CH variations. They discuss other
ways to construct \str -indices, finding a filter combination that will separate the first and second generations of globular 
cluster stars. \citet{car11} settled on 
\begin{equation}
c_y= c_1-(b-y)
\end{equation}
\begin{equation}
\delta_4= (u-v)-(b-y),
\end{equation}
with $c_y$ being defined by \citet{yon08}, and is an index which is sensitive to gravity and $N$. Both of these indices have limited use in our study, since they require
high precision photometry at the \str -$u-$ and $v-$bands. \citet{car11} point out that $m_1$ and $c_y$ have a ``complicated," degenerate dependence on metallicity (involving $[Fe/H]$ and $N$), and show that $\delta_4$ is much more
effective at estimating the $N-$abundance, and remains $CNO$-sensitive over a much broader range of stellar temperatures,
metallicities and surface gravities, since the temperature dependence is weak. \citet{car11} are more concerned with separating the N-poor, Na-poor, O-rich first generation globular population, from the N-rich, Na-rich, O-poor, second generation stars (if present). We note that there is a particular problem which involves the carbon-rich stars in the dSphs, because their colours always make a metal-poor star mimic those of a much more metal-rich object.

\section{Observations}

 As in \citetalias{hug08}, we observed the same central field (see Table 1) in \boo\ ($RA=14^h 00^m 06^s, Dec= 14.5^\circ$ J2000) with the  Apache Point
 Observatory's 3.5-m telescope, using   the direct imaging SPIcam system. The detector is a backside-illuminated SITe
 TK2048E $2048 \times 2048$ pixel CCD with 24 micron pixels, which we binned ($2\times 2$), giving a plate scale of 0.28 arc seconds per pixel,
 and a field of view (FOV) of $4.78 \times 4.78$ square arcminutes. The \citetalias{hug08} data set for  \boo\ was taken on 2007 March 19 (with a
 comparison field in M92 taken on 2007 May 24). We took 21 frames in Washington C, and Cousins R
 and I filters, with exposure time ranging from 1 seconds to 1000 seconds. The readout noise
 was 5.7e- with a gain of 3.4 e-/ADU.  The images were
 flat-fielded using dome or night-sky flats, along with with a sequence of zeros. We then
 processed the frames using the image-processing software in IRAF.\footnote{IRAF is distributed by the National Optical Astronomy Observatory, which is operated by the Association of Universities for Research in Astronomy (AURA) under cooperative agreement with the National Science Foundation.}

 The $vby$-observations used in this paper are detailed in Table 1, along with the Washington filter data from \citetalias{hug08} (when the seeing, and most airmass-values, were noticeably better). The \str\ data was taken on 2009 January 17-18, 2009 May 1, and 2011 April 5. 
The January 2009 data  used the $2\times 2 \; in^2$ \str\ filter set, which had vignetted the images.  
 The  3-inch square $uvby$ filters arrived from the manufacturer (Custom Scientific, Inc., of Pheonix, AZ) after the January 2009 observing run. We compared  the \boo\ stars observed in January and May 2009
 and found that there was no appreciable difference in the instrumental magnitudes at the same airmass. The photometric quality of the January data was better than the May data, but the January 2009 images were taken with the smaller filters. After some questions about recorded exposure times in the image headers were
 resolved, more frames were taken in April, 2011, to ensure stability of the zero-points. We also took some additional images in June, 2012 in $C$ and $R$, but the seeing was never better than $1.5\arcsec $ so they are not included.  The final weighted mean-magnitude program rejected the latter observations because of poor image quality compared to the earlier frames.
 In addition to the \boo\ central field chosen in 2007, we also observed two RGB stars, in separate fields, which had high resolution spectra in the literature: Boo-1137 and Boo-127 \citep{nor10a,nor10b,fre10,fel09}.

Table 1 lists the data taken at APO. The images taken on 2007 March 19 had sub-arcsecond seeing, and enabled us to make the best possible master source list for DAOPHOT.
For conversion to the standard Washington system, we used the \citet{gei96} Washington standard frames, containing at least 5 stars in each frame, for at least 30 standards per half-night (the APO 3.5-m is scheduled in that manner). For the \str\ data, this was more of an issue, since the \str\ system was calibrated with single stars. To reduce observing overheads, we used M92 as a cluster
standard. We used the M92 fiducial lines to assist in matching the APO data to the standard system. To supplement the \citetalias{hug08} Washington data, we took $C$ and $R$ images of the \boo\ central field in 2009 (not $I$), to make sure that there was no calibration issue with the earlier data. No problems were detected.

Employing the same data reduction method as \citetalias{hug08}, we used two iterations of (DAOPHOT-PHOT-ALLSTAR), with the first iteration having
 a detection threshold of 4$\sigma$, and the second pass had a 5$\sigma$ detection limit. We used $\sim 10$ stars in each frame to
 construct the point spread functions (PSFs), and assume that it do not vary over the chip. The chip had been found to be very stable and
 there has been no evidence that the PSF varies over the image.
 ALLSTAR \citep{ste87} was further constrained to only detect objects with a  CHI-value $< 2.0$, and almost all sources had CHI
 (the DAOPHOT goodness-of-fit statistic) between 0.5
 and 1.5 (to remove cosmic rays and non-stellar, extended objects).
 We found the aperture correction between the small (4 pixel) aperture used
 by ALLSTAR in the \boo\ field, and the larger (10-15 pixel) aperture used for the
 standards, by using the best point spread function (PSF) stars in each images.
 We used the \small IMMATCH \rm task to match the sources in each image, making several datasets in each filter combination.
 We then put together the final
 source list as follows: requiring that each star be detected in at east one image in each filter, and  the final magnitude
 and colours were calculated as the weighted (airmass, FWHM, DAOPHOT uncertainty) mean of each individual detection. 
 
 We later tested the final photometry using the standalone version of ALLFRAME \citep{ste94}, and found that the results were consistent with our method. When we used DAOPHOT III/ALLFRAME, this method produced almost identical results to those obtained by manually shifting all the images to the same positions, median-filtering them all and producing a source list from that. When we used that master-list to feed into the IRAF version of ALLSTAR, it produced 166 total detections seen at $vbyCT_1T_2(RI)$,  but the v-band data is noticeably noisier, as expected. Compared to the \citetalias{hug08} list, 117 objects were detected, and we use this group of objects in the comparison to M92.  
We find that there are 34 objects having a $v$-band uncertainty less than 0.10 dex, which were detected in multiple frames in each band (on more than one night), and had a lower overall standard-deviation-of-the-mean uncertainty; these objects are listed in Table~3. However, we plot the full-dataset in Figure 3, to show the uncertainty in magnitudes and colors, for comparison with models in \S 6.
 
In order to display how well DAOPHOT (both versions) worked on the central \boo\ field, we used the median-filtered images in each filter to show the uncertainty for each object against $T_1$ and $V(y)$. This is the best measure of how well DAOPHOT is working, and it is correlated with airmass and seeing. The images are not crowded, we have a factor of 10 fewer objets than we would detect in M92. With the artificial star experiments, the completeness of the data set is controlled by the $v$-band magnitude. The only completeness issue involves the 2 bright foreground stars seen in Figure 1c, but the source density is too low for many objects to be missed. At this point, we are not constructing luminosity functions below the MSTO, so completeness is less of a concern than the photometric uncertainties for each star. Using M92 as a cluster standard, we reduced the frames using IRAF's DAOPHOT, with 20-30 stars to fit the PSF. We then selected stars on the outer parts
of the globular cluster for testing; our photometry  yielded matches to the standard system used for $\sim 20$ randomly selected stars in common with the Frank Grundahl M92's  data set \citetext{private communication, F. Grundahl; \citealt{gru00}} of
$\sigma_{rms}=0.026$ in $V(y)$, $\sigma_{rms}=0.035$ in $(b-y)$, and $\sigma_{rms}=0.046$ in $m_1$. Our confirmation frames from 2011 were only deep enough to detect the bright RGB stars in \boo , so those stars have more observations and hence lower uncertainties in $vbyCT_1$. In Figure 4, we show colour-magnitude diagrams  used for calibration of Boo I to M92 \citep[cyan points:][]{gru00}.
The dark blue line is the Dartmouth isochrone \citep{dot08} which fits well with a recent study by \citet{dic10}, $DM=14.74$, $[Fe/H]= -2.32$, $[\alpha /Fe]=0.3$ and $Y =0.248$, and Age$=11\pm1.5$~Gyr.  \boo\ is taken to have $E(B-V)=0.02$ and $DM=19.11$, as used in \citetalias{hug08}.
 
\begin{table*}
\begin{minipage}{90mm}
\caption{APO 3.5-m CCD Frames taken in 2007-2011}
\begin{tabular}{@{}llcccc}
\hline 
{Field} & {UT$^1$}& {Filter}   & 
 {$\tau (s)$}  &    {Airmass$^2$}&   {FWHM($^{\prime\prime}$)$^3$}\\
\hline 
Boo-127& 09-01-18& y& 300& 1.19& 1.0\\ 
Boo-127& 09-01-18& b& 600& 1.17& 1.8\\ 
Boo-127& 09-01-18& v& 1200& 1.14& 1.6\\ 
Boo-1137& 09-01-18& y& 300& 1.10& 1.0\\ 
Boo-1137& 09-01-18& b& 600& 1.09& 1.3\\ 
Boo-1137& 09-01-18& v& 1200& 1.08& 0.9\\ 
Boo Ic$^4$&09-05-01& b& 1500& 1.25& 0.9\\
Boo Ic & 09-05-01& y&900& 1.33& 1.0\\
Boo Ic & 09-05-01& y&600& 1.87& 1.0\\
Boo Ic & 09-05-01& b&900& 1.65& 1.0\\
Boo Ic & 09-05-01& b&900& 2.06& 1.2\\
Boo Ic & 09-05-01& v&1200& 2.35& 1.3\\
Boo Ic & 09-05-01& v&2100& 1.48& 1.0\\
Boo Ic & 11-04-05& R&300& 1.08& 0.9\\
Boo Ic & 11-04-05& C&900& 1.07& 1.3\\
Boo Ic & 11-04-05& y&600& 1.06& 1.1\\
Boo Ic & 11-04-05& b&600& 1.06& 1.1\\
Boo Ic & 11-04-05& v&1200& 1.05& 1.3\\
Boo-1137& 11-04-05& R& 300& 1.08& 1.3\\ 
Boo-1137& 11-04-05& C& 900& 1.09& 1.4\\ 
Boo-1137& 11-04-05& y& 600& 1.11& 1.5\\ 
Boo-1137& 11-04-05& b& 800& 1.13& 0.9\\ 
Boo-1137& 11-04-05& v& 1200& 1.16& 1.0\\ 
Boo-127& 11-04-05& R& 300& 1.22& 1.1\\ 
Boo-127& 11-04-05& C& 900& 1.24& 1.5\\ 
Boo-127& 11-04-05& y& 600& 1.30& 1.3\\ 
Boo-127& 11-04-05& b& 800& 1.34& 1.2\\ 
Boo-127& 11-04-05& v& 1200& 1.40& 1.7\\ 
Boo Ic& 07-03-19 & R& 1& 1.07& 0.9\\
Boo Ic& 07-03-19 & R& 3& 1.07& 0.8\\
Boo Ic& 07-03-19 & R& 10& 1.06& 0.8\\
Boo Ic& 07-03-19 & R& 30& 1.06& 0.8\\
Boo Ic& 07-03-19 & R& 90& 1.06& 0.8\\
Boo Ic& 07-03-19 & R& 300& 1.06& 0.7\\
Boo Ic& 07-03-19 & R& 1000& 1.06& 0.8\\
Boo Ic& 07-03-19 & I& 1& 1.05& 0.6\\
Boo Ic& 07-03-19 & I& 3& 1.05& 0.6\\
Boo Ic& 07-03-19 & I& 10& 1.05& 0.6\\
Boo Ic& 07-03-19 & I& 30& 1.05& 0.6\\
Boo Ic& 07-03-19 & I& 90& 1.05& 0.7\\
Boo Ic& 07-03-19 & I& 300& 1.05& 0.7\\
Boo Ic& 07-03-19 & I& 1000& 1.05& 0.8\\
Boo Ic& 07-03-19 & C& 1& 1.06& 0.7\\
Boo Ic& 07-03-19 & C& 3& 1.06& 0.9\\
Boo Ic& 07-03-19 & C& 10& 1.06& 0.8\\
Boo Ic& 07-03-19 & C& 30& 1.06& 0.7\\
Boo Ic& 07-03-19 & C& 90& 1.06& 0.8\\
Boo Ic& 07-03-19 & C& 300& 1.07& 0.7\\
Boo Ic& 07-03-19 & C& 1000& 1.07& 0.7\\
\hline
\end{tabular}\\
(1) Year-Month-Day\\
(2) Effective airmass\\
(3) Average seeing\\
(4) Boo Ic:- \boo\ central field, see Figure 1c.\\
\end{minipage}
\end{table*}

\begin{table*} 
\begin{minipage}{84mm}
\caption{Filters Considered}
\begin{tabular}{@{}lllllr}
\hline 
Filter$^1$   &  $\lambda$(\AA ) &
      $\Delta \lambda$(\AA ) &{\scriptsize System}& {\scriptsize HST/WFC3$^2$} & {$\tau(s)^3$} \\
      \hline
u& 3520& 314& \str\ & F390M & 18,980 \\
 v &  4100 & 170& Str\"{o}mgren & F410M& 10,214 \\
b &  4688 & 185& Str\"{o}mgren & F461M& 5318 \\
 y &  5480 & 226& Str\"{o}mgren & F547M & 1,494 \\
 C & 3980& 1100& Washington & F390W & 2,612 \\
$T_1$ &  6389& 770& Use $R_C$& F625W & 605 \\
 $T_2$ & 8051& 1420& Use $I_C$ & F775W & 390 \\
$R$ &  6407& 1580& Use $R_C$& F625W & 605 \\
 $I$ & 7980& 1540& Use $I_C$ & F775W & 390 \\
 u &  3596& 570& SDSS & F336W & 6,336 \\
 g & 4639& 1280& SDSS & F475W &  835 \\
 r & 6122& 1150& SDSS & F625W &  605\\
 i &  7439& 1230& SDSS & F775W & 988 \\
\hline
\end{tabular}\\
(1) Filter data from \citet{bes05}.\\
(2) HST/WFC3 filter best equivalent to ground-based choice.\\
(3) Estimated exposure time for a G2V star at $V \sim 23$ mag. at the distance of \boo\ for $S/N=50$.
\end{minipage}
 \end{table*}

 We solved for each filter, rather than  the \str\ indices, for each night. The transformation equations on 2009 May 1 are as follows:
\begin{equation}
V=y_i-2.187-0.012(b_i-y_i)-0.163X, \; \sigma_{rms}=0.007
\end{equation}
\begin{equation}
b=b_i -2.217+0.031(b_i-y_i)+0.220X, \; \sigma_{rms}=0.017
\end{equation}
\begin{equation}
v=v_i-2.464-0.512(v_i-b_i)-0.300X, \; \sigma_{rms}=0.020
\end{equation}
Here, \it X \rm denotes the effective airmass and the subscript \it i \rm indicates the instrumental magnitude. The $\sigma_{rms}$ values are the comparison with the standards. 

\citetalias{hug08}'s photometry from 2007 March 19 yielded matches to the \citetalias{gei99} standard system of
$\sigma_{rms}=0.021$ in $T_1$, $\sigma_{rms}=0.015$ in $(C-T_1)$, and $\sigma_{rms}=0.017$ in $(T_1-T_2)$. 
In $T_1$, the average uncertainties in the final CMD were $\sigma_{rms}=0.024$ at the level of the horizontal branch, 
and $\sigma_{rms}=0.04$ at the MSTO. The transformation equations are as follows:
\begin{equation}
T_1=R_i-0.461+0.021(C_i-R_i)-0.150X, \; \sigma_{rms}=0.021
\end{equation}
\begin{equation}
(C-T_1)=1.117(C_i-R_i)-1.015-0.322X, \; \sigma_{rms}=0.015
\end{equation}
\begin{equation}
(T_1-T_2)=1.058(R_i-I_i)+0.460-0.046X, \; \sigma_{rms}=0.017
\end{equation}
As before, \it X \rm denotes the airmass and the subscript \it i \rm indicates the instrumental magnitude.

 The final calculated uncertainties for each individual star in
 each image were found  by taking the uncertainties from photon statistics, DAOPHOT's uncertainties, the aperture corrections, and the standard photometric errors in quadrature. We then took the weighted means of multiple observations, which
 reduced uncertainties internal to the data set. Thus, we achieved better uncertainties in for each star by taking the weighted means, with the weighting being dependent on the DAOPHOT uncertainties and the airmass, which was found to be equivalent to the seeing. We constructed image sets comparing short and long exposures, keeping to similar airmass and seeing between frames, to achieve multiple, independent observations of each star and improved the final (standard-deviation-of-the-mean) uncertainty for the objects above $T_1\sim 22 ~mag.$
 We were able to detect 166 objects in the field in the $vbyCT_1T_2(RI)$-filters.  A
 finding chart for the objects in Table~3 (117 objects) is given in \citetalias{hug08}, and is shown in Figure 1b and 1c here.
 Also, from Table~3, the 19 brightest stars were detected in the SDSS survey\footnote{accessed through http://www.sdss.org/ DR5 and DR6 \citep{ade08}}, and we include them in Table~4.
 
 An independent, external test on how well we have calibrated the data is discussed in \S 6.1, which covers spectral energy distributions (SEDs).

 \section{STATISTICAL REMOVAL OF NON-MEMBERS}

The process used by \citetalias{hug08} to remove non-dSph stars from our final data set was modified from that used on the globular clusters, NGC 6388 and $\omega$~Cen  \citep[][respectively]{hug07,hug00}. Due to observing-time constraints, we did not observe an off-galaxy field, but instead generated artificial field stars with the TRILEGAL code \citep{gir05}, calibrated for the SPIcam FOV and the appropriate magnitude limits. Briefly, we can statistically compare CMD of the off-galaxy region (or the simulated field) to the \boo\ field. The method was adapted from \citet{mig98}. 
Here, we detail the method used to remove foreground objects from the \citetalias{hug08} data set, which 
 was also used with the \str\ for consistency, yielding similar results. The meaning of the letter grades in Table 3 is as follows. Class A are sources which have passed statistical cleaning and colour-selection, and which have uncertainties better than 0.05 in all Washington filters. Class B are sources which have passed statistical cleaning and colour-selection, which do not have uncertainties better than 0.05 in all filters. Class C are sources which passed colour-selection failed statistical cleaning, and which have uncertainties better than 0.05 in all filters. Class D objects passed colour-selection, failed statistical cleaning, and do not have uncertainties better than 0.05 in all filters. Class E sources passed statistical cleaning but failed colour selection, and Class F failed statistical cleaning and colour selection. 
Where the whole image is filled by the target (galaxy or globular cluster), we would have to use an off-target field (or simulate one). For each star in the on-target image sub-section (or separate image), we count the number of stars in the colour-magnitude diagram that have  $C$-magnitudes \citepalias{hug08} within max (2.0, 0.2) mag. of the $C$ and $C-T_1$  colours of the supposed dwarf-population stars in the CMD. We call this number $N_{on}$. Now, we also count the number of field stars, in the off-dwarf image or image sub-section (or simulated field), that fall within the same ranges in the CMD, and call this number $N_{off}$.

We calculate the probability that the star in the on-dwarf field CMD is a member of the dwarf galaxy population as: 
\begin{equation}
p\approx 1-min\left({\alpha N_{off}^{UL \; 84}\over N_{on}^{LL\; 95}},\; 1.0\right)
\end{equation}
			                  
Where $\alpha$ is the ratio of the area of the dSph galaxy region to the area of the (simulated) field region and
\begin{multline}
N_{off}^{UL \; 84}\approx \\
(N_{off}+1)\left[  1- {1\over 9(N_{off}+1)}
+{1.000\over 3 \sqrt{N_{off}+1}}\right ]^3
\end{multline}
The equations are taken from the Appendix of \citet{hug00}, and 
corresponding to eq. [2] of \citet{mig98} and eq. [9] of \citet{geh86}. Here, eq.[14] is the 
estimated upper (84 per cent ) confidence limit of $N_{off}$, using Gaussian statistics. 
\begin{multline}
N_{on}^{LL\; 95}\approx \\
N_{on}\times 
\left[1-{1\over 9N_{on}} - {1.645\over 3 \sqrt{N_{on}}}
+0.031N_{on}^{-2.50}\right]^3
\end{multline}

Then,  eq.[15]  is then the lower 95 per cent  confidence limit for $N_{on}$ (eq. [3] of \citet{mig98}, and eq. [14] of \citet{geh86}). For a large, relatively nearby cluster like $\omega$~Cen, we assumed that the whole on-cluster field is part of the system (a fairly safe assumption), so that $\alpha$ is assumed to be 1, in that case. Here, we  generated a population of MWG stars with the TRILEGAL code for the same sky area as the SPIcam FOV, so that $\alpha = 1$ for \boo , also.
Then, in order to estimate if any particular star is a cluster/dwarf member, we generate a uniform random number, $0<p^\prime<1$, and if (eq.[13]'s) $p>p^\prime$, we accept the star as a member of the cluster or dwarf galaxy. This method works best if there is a colour/metallicity difference between the foreground and \boo\ dwarf populations, which means it is less effective at 
removing field stars if we use the SDSS-filters.

Figure 3a--c shows the final photometric uncertainties from the Washington filter data. We re-reduced the data, and made a median-filtered images for each $vbyCT_1T_2(RI)$-filter, shifted the images, and made a master list of objected detected in a summed master-median-filtered image. The open circles are the 166 objects detected in all the median-filters images. Figure 3d--f shows the uncertainty distribution for the 166 detections in the $vby$-filters as open circles.  The Washington-filter images had fainter magnitude limits than the \str\ images, so we continue to use  the statistical cleaning results from \citetalias{hug08}, but we can use the \str\ indices to estimate photometric metallicities and compare with the statistical results. Observing time in the v-filter is controlling our limiting magnitude.
We compared the 166 objects detected in the median-filtered images in all 6 bands, compared them with the 165 objects from \citetalias{hug08}, and 117 objects passed the DAOPHOT CHI-value $< 2.0$ in the $v$-band, and we merged the two lists. Table~3  contains the 34 objects which were observed on more than one night in all filters and had uncertainties $< 0.1$ in the $v$-band. The analysis from this point onwards requires that all objects have at least $vbyCT_1$-magnitudes.

\begin{table*} 
\centering
\begin{minipage}{176mm}
\caption{Objects in \boo\ with Washington \& \str\ Photometry}
\begin{tabular}{@{}llllllllrlll}
\hline 
{ID$^1$}& {$X_R^2$}& {$Y_R$}& RA& DEC& {Class$^3$}& {$T_1$}&  {$(C-T_1)$}&  {$(T_1-T_2)$}&
 {$V$}&  {$(b-y)$}&  {$m_1$}\\
 \hline 
\rowcolor{LC}
Boo-1137&	$-$&	$-$&	13:58:33.82&	14:21:08.5&	A & 17.08(0.02)&	1.69(0.03)&	0.61(0.02)&	17.65(0.01)&	0.62(0.02)&	0.09(0.03)		\\
\rowcolor{LC}
Boo-127&	$-$&	$-$&	14:00:14.57&	14:35:52.1&	A & 17.12(0.02)&	1.84(0.03)&	0.57(0.02)&	17.68(0.004)&	0.68(0.02)&	0.14(0.03)		\\
\rowcolor{LC}
Boo-117/HWB-8&	298.68&	891.95&	14:00:10.49&	14:31:45.6&	C &17.20(0.02)&	1.82(0.03)&	0.61(0.02)&	17.79(0.004)&	0.61(0.02)&	0.16(0.03)	\\
\rowcolor{LC}
Boo-119/HWB-9&	330.67&	171.27&	14:00:09.85&	14:28:23.1&	A & 17.48(0.02)&	1.80(0.03)&	0.57(0.02)&	17.98(0.005)&	0.63(0.02)&	0.20(0.03)		\\
\rowcolor{LC}
HWB-22&	950.94&	97.89&		13:59:57.85&	14:28:02.8&	A & 19.77(0.02)&	1.16(0.04)&	0.47(0.02)&	20.22(0.01)&	0.47(0.02)&	0.04(0.04)	\\
\rowcolor{LC}
HWB-24&	667.94&	272.14&	14:00:03.33&	14:28:51.6&	C & 20.10(0.02)&	1.06(0.03)&	0.50(0.02)&	20.48(0.01)&	0.42(0.03)&	0.04(0.06)		\\
\rowcolor{LC}
HWB-28&	564.80&	599.22&	14:00:05.34&	14:30:23.5&	A & 20.40(0.02)&	1.16(0.04)&	0.50(0.03)&	20.88(0.02)&	0.45(0.03)&	0.09(0.04)		\\
\rowcolor{LC}
HWB-34&	681.50&	600.22& 14:00:03.08&	14:30:23.8&	A &	20.86(0.02)&	1.09(0.04)&	0.46(0.02)&	21.27(0.02)&	0.44(0.04)&	0.03(0.06)		\\
HWB-3& 960.10& 499.42& 13:59:57.69& 14:29:55.6& F& 16.01(0.01)& 3.47(0.03)& 1.65(0.03)& 16.91(0.03)& 1.27(0.03)& -0.03(0.04)\\
HWB-4& 54.57& 53.07& 14:00:15.18& 14:27:49.8& F& 16.24(0.01)& 3.29(0.03)& 1.28(0.03)& 17.17(0.02)& 1.02(0.03)& 0.33(0.05)\\
HWB-6& 630.81& 818.25& 14:00:04.07& 14:31:25.1& E& 16.86(0.01)& 1.20(0.03)& 0.49(0.02)& 17.26(0.02)& 0.44(0.03)& 0.08(0.05)\\
HWB-11& 891.80& 840.98& 13:59:59.02& 14:31:31.5& E& 17.71(0.01)& 3.26(0.03)& 1.10(0.02)& 18.67(0.01)& 0.97(0.02)& 0.46(0.04)\\
HWB-14& 540.96& 301.43& 14:00:05.78& 14:28:59.8& E& 18.63(0.01)& 0.96(0.03)& 0.40(0.02)& 18.95(0.02)& 0.40(0.03)& 0.09(0.05)\\
HWB-15& 461.31& 912.01& 14:00:07.35& 14:31:51.3& E& 18.79(0.01)& 3.39(0.03)& 1.30(0.02)& 19.82(0.02)& 1.04(0.03)& 0.37(0.06)\\
HWB-16& 287.26& 389.96& 14:00:10.69& 14:29:24.6& A& 18.92(0.01)& 1.37(0.03)& 0.52(0.02)& 19.39(0.02)& 0.52(0.03)& 0.07(0.04)\\
HWB-17& 72.98& 337.39& 14:00:14.84& 14:29:09.7& F& 19.06(0.01)& 0.88(0.03)& 0.38(0.02)& 19.40(0.01)& 0.35(0.02)& 0.08(0.04)\\
HWB-18& 171.21& 684.41& 14:00:12.95& 14:30:47.3& E& 19.23(0.01)& 2.11(0.03)& 0.59(0.02)& 19.83(0.01)& 0.59(0.02)& 0.46(0.04)\\
HWB-19& 626.80& 6.28& 14:00:04.11& 14:27:36.9& A& 19.33(0.02)& 0.87(0.03)& 0.38(0.02)& 19.59(0.01)& 0.36(0.03)& 0.07(0.04)\\
HWB-20& 329.81& 701.76& 14:00:09.88& 14:30:52.2& E& 19.37(0.01)& 2.73(0.03)& 0.86(0.02)& 20.18(0.02)& 0.80(0.04)& 0.51(0.06)\\
HWB-21& 963.85& 2.41& 13:59:57.59& 14:27:35.9& E& 19.57(0.01)& 3.12(0.06)& 0.94(0.02)& 20.38(0.02)& 0.96(0.03)& 0.32(0.09)\\
HWB-26& 589.61& 886.86& 14:00:04.87& 14:31:44.3& E& 20.30(0.02)& 2.78(0.04)& 0.88(0.02)& 21.13(0.02)& 0.77(0.04)& 0.59(0.07)\\
HWB-29& 510.88& 714.46& 14:00:06.38& 14:30:55.8& A& 20.47(0.01)& 1.20(0.03)& 0.63(0.03)& 20.91(0.02)& 0.46(0.03)& 0.08(0.07)\\
HWB-31& 207.93& 381.42& 14:00:12.23& 14:29:22.1& C& 20.72(0.01)& 0.97(0.03)& 0.46(0.03)& 21.09(0.02)& 0.44(0.03)& -0.04(0.04)\\
HWB-32& 692.68& 332.82& 14:00:02.85& 14:29:08.7& E& 20.77(0.01)& 0.76(0.03)& 0.36(0.02)& 20.99(0.03)& 0.32(0.03)& 0.03(0.05)\\
HWB-33& 848.41& 14.94& 13:59:59.83& 14:27:39.4& E& 20.84(0.01)& 0.96(0.04)& 0.36(0.03)& 21.14(0.02)& 0.44(0.03)& -0.04(0.05)\\
HWB-36& 653.27& 903.42& 14:00:03.64& 14:31:49.0& E& 20.95(0.02)& 1.12(0.04)& 0.41(0.03)& 21.40(0.03)& 0.44(0.04)& 0.01(0.06)\\
HWB-37& 925.81& 867.63& 13:59:58.36& 14:31:39.0& E& 21.06(0.01)& 1.80(0.05)& 0.59(0.02)& 21.60(0.03)& 0.52(0.05)& 0.31(0.08)\\
HWB-40& 789.33& 824.54& 14:00:01.00& 14:31:26.9& A& 21.28(0.02)& 1.13(0.04)& 0.50(0.03)& 21.64(0.03)& 0.46(0.04)& 0.05(0.07)\\
HWB-44& 945.51& 434.38& 13:59:57.97& 14:29:37.3& E& 21.52(0.02)& 1.00(0.04)& 0.39(0.03)& 21.86(0.04)& 0.43(0.05)& -0.06(0.08)\\
HWB-45& 545.05& 475.10& 14:00:05.71& 14:29:48.6& A& 21.54(0.01)& 0.21(0.03)& 0.57(0.03)& 22.03(0.04)& 0.34(0.05)& -1.02(0.07)\\
HWB-47& 673.78& 935.68& 14:00:03.24& 14:31:58.1& A& 21.75(0.02)& 0.92(0.04)& 0.47(0.04)& 22.11(0.04)& 0.36(0.06)& 0.10(0.09)\\
HWB-48& 845.80& 760.07& 13:59:59.91& 14:31:08.8& A& 21.82(0.02)& -0.08(0.04)& 0.17(0.05)& 21.78(0.03)& 0.07(0.05)& 0.08(0.06)\\
HWB-50& 87.95& 538.21& 14:00:14.56& 14:30:06.1& A& 21.92(0.02)& 0.86(0.05)& 0.40(0.03)& 22.28(0.05)& 0.38(0.07)& 0.02(0.09)\\
HWB-51& 661.84& 801.67& 14:00:03.47& 14:31:20.4& A& 21.92(0.02)& 0.33(0.04)& 0.28(0.05)& 22.05(0.05)& 0.24(0.06)& 0.05(0.08)\\\hline
\end{tabular}\\
(1) $ID$ from \citetalias{hug08}, proper motion-confirmed members listed first.\\
(2) Positions from the Figure 1b.\\
(3) \citetalias{hug08}'s object classes:\\
\hskip 0.5cm A - If sources passed the statistical cleaning process, had the correct colours and had photometry in all filters with uncertainties less than 0.05.\\
\hskip 0.5cm B -  Objects passed the cleaning program, but had uncertainties in all filters not less than 0.05.\\
\hskip 0.5cm C - Passed the statistical cleaning process, had the correct colours and A-type good photometry, but failed the comparison with the randomly generated probability.\\
\hskip 0.5cm D - Objects failed the statistical cleaning process (also had the right colours but poor photometry).\\
\hskip 0.5cm E -  Passed statistical cleaning but failed colour selection \citepalias[according to][]{hug08}.\\
\hskip 0.5cm F -  These objects failed both statistical cleaning and colour selection, and tended to be well outside the CMD area of a metal-poor dwarf. Usually bright foreground stars.\\
\end{minipage}
 \end{table*}

\begin{figure} 
\includegraphics[scale=1.0,width=74mm]{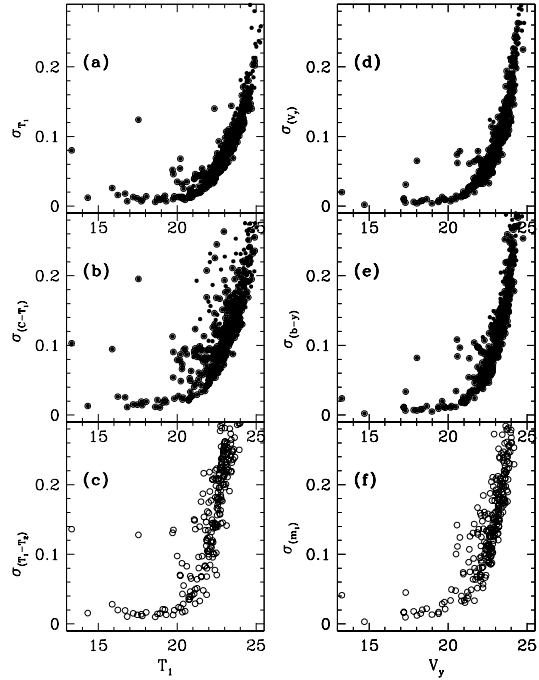}
 \caption{
 Uncertainty vs. magnitude plots for the Boo I data sets, from \citetalias{hug08} and this paper. Open circles are 166 stars with $CT_1T_2$-measurements. These are the DAOPHOT uncertainties from the master-median-filtered images in each of the 6 filters.
\bf (a) \rm Uncertainty in $T_1$ vs. $T_1$. 
\bf (b) \rm Uncertainty in $(C-T_1)$ vs. $T_1$.  
\bf (c) \rm Uncertainty in $(T_1-T_2)$ vs. $T_1$. 
\bf (d) \rm Uncertainty in $V_y$ vs. $V_y$, which is Johnson-$V$ calculated from Str\"{o}mgren-$y$. \bf (e) \rm Uncertainty in $(b-y)$ vs. $V_y$. 
\bf (f) \rm Uncertainty in $m_1$ vs. $V_y$. }
\end{figure}

\section{Photometric Metallicities} 

Following on from the discussion in \S 2,  several authors have 
generated photometric calibrations on the \str\ [Fe/H]-scale, amongst them, \citet{hil00} and \citet{cal07}. In this section, we select several calibrations from those papers, where $m_0$ is the dereddened $m_1$-index and $[m]$ is the reddening-free version:

\begin{multline}
Hilker \; (2000)\; :\; \\
[Fe/H]_{Hil}={m_0-1.277(b-y)_0+0.331\over 0.324(b-y)_0-0.031}.
\end{multline}
\begin{multline}
Calamida \; et\;  al.\; (2007):\; \\
[Fe/H]_{m_1}={m_0+0.309-0.521(v-y)_0\over 0.159(v-y)_0-0.090}.
\end{multline}
\begin{multline}
Calamida \; et\;  al.\; (2007):\; \\
[Fe/H]_{[m]}={[m]+0.251-0.585(v-y)_0\over 0.131(v-y)_0-0.070}.
\end{multline}
These methods of calculating photometric metallicities are used for columns 2--4 of Table 5.

\begin{figure} 
\vskip 2cm
\includegraphics[width=84mm,scale=1.0]{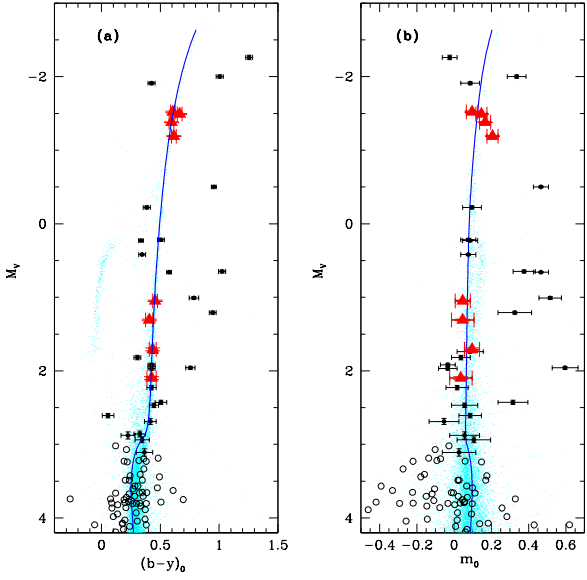}
\caption{\bf (a) \rm $M_V \; vs. \; (b-y)_0$ Str\"{o}mgren CMD for M92 (cyan points; data provided by F. Grundahl), Boo I Str\"{o}mgren data only (black points),
Str\"{o}mgren and Washington objects (black filled circles), and proper-motion members with
Str\"{o}mgren, Washington and SDSS magnitudes (red filled triangles). 
The dark blue line is the Dartmouth isochrone corresponding to $[Fe/H]=-2.25$, $[\alpha /Fe]=0.3$ and an age of 11~Gyr. M92 has $E(B-V)=0.025$ and $DM=14.74$.  Boo I has $E(B-V)=0.02$ and $DM=19.11$. \bf (b) \rm  $M_V \; vs. \; m_0$, the Str\"{o}mgren CMD for M92 (cyan points), Boo I Str\"{o}mgren data only (black points), Str\"{o}mgren and Washington objects (black filled circles), and proper-motion members with Str\"{o}mgren, Washington and SDSS magnitudes (red filled triangles). The dark blue line is the Dartmouth isochrone corresponding to $[Fe/H]=-2.25$, $[ \alpha /Fe]=+0.3$, and an age of 11.0~Gyr.}
\end{figure}

In Figure 5a and 5b,  we show colour-colour plots and [Fe/H]-calibrations for M92 (cyan points). The blue points  are the M92 RGB stars above the horizontal branch (HB). Having the same type of cool, metal-poor RGB as the expected dSph 
population, these plots illustrate the loss of metallicity resolution on the lower-RGB in the
Str\"{o}mgren system.  We used the TRILEGAL code\footnote{http://stev.oapd.inaf.it/cgi-bin/trilegal \citep{van09}}  to generate a field of artificial stars at the correct galactic latitude, for the same magnitude limits as our dSph field.  Figure 5c and 5d show our \boo\ data from Table~3 (black points with error bars) and the TRILEGAL-generated artificial stars (blue circles). The red triangles are the bright RGB stars with SDSS-colours. The Str\"{o}mgren filters are well-suited to separate the dSph population from the
foreground stars. We note that the TRILEGAL artificial stars have the same colors as the foreground RGB stars, and are separate from the upper-RGB proper-motion members Figure 5c and 5d; all data points overlap for stars which likely have $\log ~g>2.5$, which is not a function of photometric uncertainty, it is the \str\ bands losing sensitivity to metallicity. We confirm that the \str\  indices are only useful for upper RGB stars from these plots alone.

\begin{figure} 
\includegraphics[width=84mm,scale=1.0]{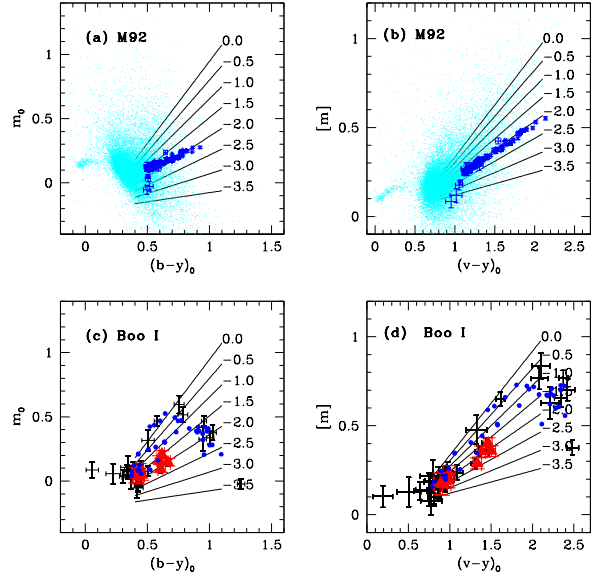}
\caption{
\bf (a) \rm Plot of $m_0=(v-b)_0-(b-y)_0$ vs. $(b-y)_0$ for M92 RGB stars (blue points, F.Grundahl, private communication), and the rest of the globular cluster's stars (cyan). The calibration lines of constant [Fe/H] are taken from \citet{hil00}. 
\bf (b) \rm For the same M92 sample, we show
 $[m]=m_1+0.3(b-y)$, the reddening-free index, plotted against $(v-y)_0$. Calibration from \citet{cal07}. 
 \bf (c)  \rm $m_0=(v-b)_0-(b-y)_0$ vs. $(b-y)_0$ for our sample (from the median filtered images), with the \citet{hil00} calibration. In total, 117 objects were detected
 in $vby$ filters, shown as black points. The TRILEGAL code was used to generate a sample
 of foreground stars, shown as blue filled circles. The  RGB proper-motion members are shown as red triangles. 
\bf (d) \rm  $[m]=m_1+0.3(b-y)\; vs. \; (v-y)_0$ for the same sample of \boo\ stars, with the \citet{cal07} calibration.}
\end{figure}

We use the usual \str\ relationships:
\begin{align}
E(b-y)&=0.70E(B-V);\\
E(v-y)&=1.33E(B-V);\\
E(m_1)&=-0.30E(b-y);\\ 
{[m]}&=m_1+0.30(b-y).
 \end{align}
The \it reddening-free \rm metallicity index is $[m]$, as used by \citet{cal07}; the reddening law is taken from \citet{car89} and \citet{cra75}.
 
\citet{cal07} generated several versions of the \str -$[Fe/H]$ calibration, which we tested with our  Table~3 data; their 
calibration equations that best fit our data are given in
equations [17] \& [18]. These calibrations were chosen because they best matched the spectroscopy available in 2010 \citep{fel09,nor08,iva09,mar07}. The \citet{cal07} equations are based on ``semiempirical" calibrations of \citet{cle04}, and have differences between the $[Fe/H]_{phot}$ and $[Fe/H]_{spec}$ values of $-0.06\pm 0.18$ dex and $-0.05 \pm 0.18$ dex, respectively. 

When \citet{cal07} tested the \citet{hil00} calibration on 73 field RGB stars, the difference between the schemes was found to be $0.13\pm 0.20$ dex. \citet{hug04} used the \citet{hil00} calibration for their study of $\omega$~Cen, which \citet{cal09} found to be in agreement with their work.
For stellar populations in dSph galaxies, it is impractical to use any calibration involving the $u$-band, both because the RGB and MS stars are too faint in the near-UV, but \str -$u$  is more sensitive to reddening than  $v$ or $C$. In practice, half of all observing time would have to be dedicated to $u$-band imaging, to have a chance of obtaining $c_1$-indices, so we do not consider these here.
Figure 5a shows an interesting characteristic of the \citet{hil00} calibration, in that the RGB (dark blue points) of M92 is skewed 
with respect to the semi-empirical lines of constant $[Fe/H]$, and the tip of the RGB would appear more metal poor than the lower part of the RGB. Figure 5b confirms that \citet{cal07}'s $[m]\; vs. \; (v-y)$ calibration fits well with M92 having $[Fe/H]\approx -2.2$. When we apply the calibrations to our data in Table 5, this skewing is observed (comparing column 2 to columns 3 \& 4). Again, we notice that the fainter RGB stars have large uncertainties, resulting from the loss of sensitivity of the $m_1$-index and the increasing photometric uncertainties, particularly at \str -$v$. 

For the Washington filters, we use:
\begin{align}
E(C-T_1) &= 1.966E(B-V);\\
E(T_1-T_2) &= 0.692E(B-V);\\
M_{T_1}&=T_1+ 0.58E(B-V) - (m-M)_V;
\end{align}
from \citet{gei91} and \citetalias{gei99}. 
\citetalias{hug08} used equations (23)--(25) and the \citetalias{gei99} standard giant branches to find the $[Fe/H]$-values for the \boo\ RGB stars.  Note that \citetalias{gei99} use $A_V = 3.2E(B-V)$; not setting $R_V=3.1$ does not transform into an appreciable  difference for \boo , as $E(B-V)=0.02$.

In Table 6, we show the preliminary results from \citet{hug11b}, where the stellar
parameters were estimated from $\chi^2$ fits to  the $[\alpha/Fe]=0.0$ to $+0.4$ Dartmouth isochrones \citep{dot08}. We found that the best fits gave $[\alpha/Fe]=+0.3 \pm 0.1$ dex. The early results presented in \citeauthor{hug11b}'s \citeyearpar{hug11b} conference paper were consistent with the \citet{mar07}, CaT-based spectroscopy, available at the time. However, the model grid was not fine enough for our goals, and we ran extensive, finer-grid models, based on the Dartmouth isochrones.

\begin{table*}
\begin{minipage}{116mm}
\caption{SDSS Magnitudes for Stars in \boo\ Central Field}
\begin{tabular}{@{}lcccccl}
\hline 
ID& SDSS $u$&  $g$&  $r$&  $i$&   $z$&
Class\\
\hline
Boo-1137& 19.77(03)& 18.11(01)& 17.37(01)& 17.04(01)& 16.86(01)& Member$^1$\\
Boo-127& 20.01(05)& 18.16(01)& 17.37(01)& 17.02(01)& 16.85(01)& Member$^1$\\
HWB-3& 20.10(05)& 17.73(01)& 16.28(01)& 14.77(01)& 13.98(01)&  F    \\
HWB-4& 20.47(05)& 18.00(01)& 16.57(00)& 15.65(01)& 15.16(01)&  F    \\
HWB-6& 23.37(74)& 22.43(12)& 20.96(04)& 19.47(02)& 18.76(04)&  E\\
Boo-117/HWB-8& 19.98(05)& 18.21(01)& 17.44(01)& 17.10(01)& 16.92(01)&  C \\
Boo-119/HWB-9& 20.21(04)& 18.43(01)& 17.69(01)& 17.38(01)& 17.21(01)& A     \\
HWB-11& 21.74(20)& 19.49(01)& 18.05(01)& 17.23(01)& 16.77(01)& E    \\
HWB-15& 24.46(13)& 20.63(03)& 19.22(01)& 18.18(01)& 17.63(02)& E    \\
HWB-16& 21.08(11)& 19.70(02)& 19.13(01)& 18.87(01)& 18.72(04)&  A   \\
HWB-18& 22.48(36)& 20.42(02)& 19.54(02)& 19.16(02)& 19.01(05)&  E   \\
HWB-20& 23.47(80)& 20.90(03)& 19.62(02)& 19.03(02)& 18.70(04)&  E\\
HWB-21& 24.50(67)& 21.22(03)& 19.86(02)& 19.18(02)& 18.77(03)&  E\\
HWB-22& 21.63(11)& 20.35(02)& 19.85(02)& 19.63(02)& 19.55(06)&  A   \\
HWB-23& 24.11(61)& 21.92(06)& 20.31(02)& 19.00(01)& 18.24(02)& E\\
HWB-24& 21.69(11)& 20.74(02)& 20.21(02)& 20.02(03)& 19.99(08)& C\\
HWB-26& 25.01(12)& 21.82(07)& 20.59(03)& 19.94(03)& 19.64(08)& E\\
HWB-28& 22.51(38)& 21.04(04)& 20.63(03)& 20.42(04)& 20.40(15)&  A\\
HWB-29& 22.24(29)& 21.00(04)& 20.69(03)& 20.50(05)& 20.27(13)&  A\\
HWB-31& 21.71(19)& 21.27(05)& 20.87(04)& 20.77(06)& 20.55(17)&  C \\
HWB-34& 22.67(43)& 21.47(05)& 21.00(05)& 20.81(06)& 20.44(15)& A\\
\hline
\end{tabular}\\
 (1) Confirmed Boo I members outside central field.
 \end{minipage}
\end{table*}

We expect that the dSph stars are $\alpha$-enhanced in the same way as the MWG halo population \citep[see discussion in][and our $\S 5$]{coh09,nor08}.  We use both the $\alpha$-enhanced and solar-scaled models from the Dartmouth Stellar Evolution Database\footnote{http://stellar.dartmouth.edu/$\sim$models/} \citep{dot08,cha01,gue92} to compare with our data.

\begin{figure} 
\vskip 2cm
\includegraphics[width=84mm,scale=1.0]{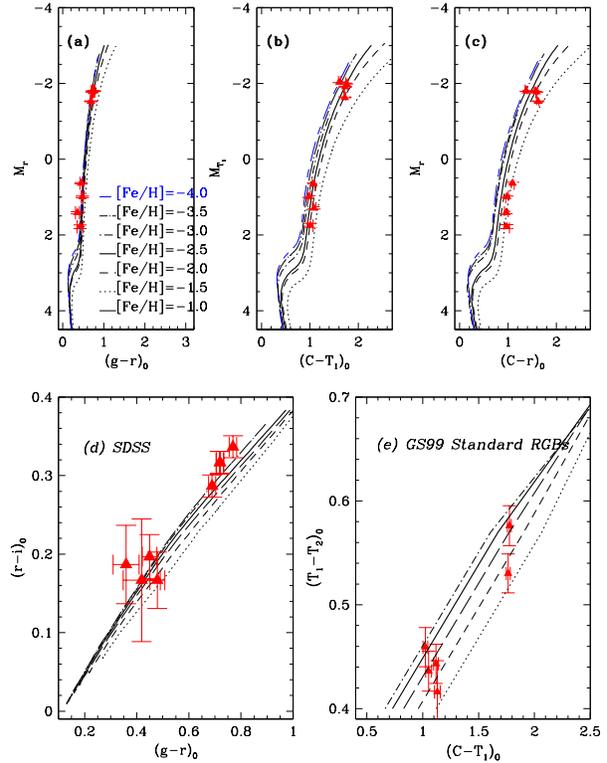}
\caption{CMDs and colour-colour plots using a mixture of SDSS and Washington filters to show the [Fe/H]-sensitivity (HWD). In all diagrams, the red triangles are the 8 RGB stars, known to be proper-motion members. All isochrones shown are solar-scaled and have  an age of 12~Gyr.
     \bf (a) \rm $M_r \; vs.\;  (g-r)_0$ with Dartmouth models, $\alpha$-enhanced to +0.3.
      \bf (b) \rm $M_{T_1} \; vs.\;  (C-T_1)_0$ with Dartmouth models, also $\alpha$-enhanced to +0.3.
      \bf (c) \rm $M_r \; vs.\;  (C-r)_0$
      \bf (d) \rm $(r-i)_0 \; vs. \; (g-r)_0$ SDSS colour-colour plot, with the Dartmouth isochrones.
 \bf (e) \rm Washington colours with \citetalias{gei99} standard giant branches \citepalias{hug08}.}
\end{figure}

\section{Discussion}

From Figure 6b, we see that $(C -T_1)$ widens the separation of the giant branches of different metallicities, giving a resolution for the \citetalias{gei99} RGB fiducial lines of $\sim 0.15$ dex. One of us (GW) suggested the Washington system, which was developed by \citet{can76,gei96} defined CCD standard fields for the system.   In
\citetalias{hug08}, we found that Washington filters spread out the stars at the MSTO, and we have found that $(C - T_1)$ is more effective than the SDSS-colours $(g - r)$, as shown in Figure 6a. The SDSS photometry is not sensitive enough to this difference in colours to distinguish Boo~I's level of metallicity spread. The  subset of 19 objects  with $ugriz$ magnitudes (Table 4) contains most of the bright objects from \citetalias{hug08} and this paper's \str\ data set. These are the objects which have  \str\ and Washington colours falling in the appropriate ranges which enable us to calculate the photometric values of $[Fe/H]$. Table~3 contains foreground stars and  \boo\ members, and intersects with the spectroscopic data set of \citet{mar07}. We have radial velocities and independent metallicities for 8 of the objects in Table~3, and \str\ and Washington $[Fe/H]_{phot}$-values for many of them.  We added the 2 giants outside the central \boo\ field, Boo-1137 and Boo-127, as additional calibrators, because Boo-1137 \citep{nor10b} is very metal poor, and both have had a number of spectra taken (see Table 5). 

We considered the effect of replacing the $g$-filter with the much broader $C$-band \citep{hug11a}. As we can see from Figure 6b and 6c, the
colour, $(C-r)$ works well for the upper RGB ($\log g < 1.5$) in \boo , but that the lower-RGB stars 
($2.5 < \log g < 3.0$) are not well-separated in this colour. The reasons for this discrepancy are filter-width and transmission, which are shown  in Figure 2a and Table 2. While the Sloan $r$-filter has a greater overall transmission than the
Washington $T_1$-filter, it has become standard practice to substitute the broadband $R$-filter for
$T_1$, since it has $\Delta \lambda_R=1580$\AA , compared to $\Delta \lambda_{T_1}=770$\AA\ \citep[][also see Table 2]{gei96}.

As with the \citet{lai11} method, there are merits in combining all the available photometry,
 but we are searching for the minimum useful number of filters that can break the age-metallicity
 degeneracy. 
In Figure 7, we compare the Str\"{o}mgren and Washington filters, and construct two new indices: 
\begin{equation}
m_*=(C-T_1)_0-(T_1-T_2)_0
\end{equation}
and 
\begin{equation}
m_{**}=(C-b)_0-(b-y)_0.
\end{equation} 

Our motivation in defining these indices is to avoid the 
collapse of the metallicity sensitivity of the $m_1$-index on the lower RGB, and to attempt 
to replace the $v$-filter with the broader $C$-filter \citep[also see][]{hug11a}. 
  Figure 7h shows $m_{**} \; vs.\;  (C-T_1)_0$, and  allows us to use 4 filters, $CT_1by$, which reduces observing time and 
keeps $\sim  0.3$~dex [Fe/H]-resolution for stars with $-1.5>[Fe/H]>-4.0$. Figure 7i shows that 
we could use $Cby$ for metallicity estimates  $-1.5>[Fe/H]>-4.0$. These colour-colour plots are not sensitive to age on the RGB, 
recovering age-resolution at the MSTO. However, reaching the MSTO with $Cby$ is much faster than with $vby$ (see Table 2). Figs.7a, b \& c show that $m_0$ may be preferred for systems with $-1.0>[Fe/H]>-2.0$, but we note that $(C-T_1)$ would also be more useful there, with much shorter exposure times.

\begin{figure} 
\vskip 2cm
\includegraphics[width=84mm,scale=1.0]{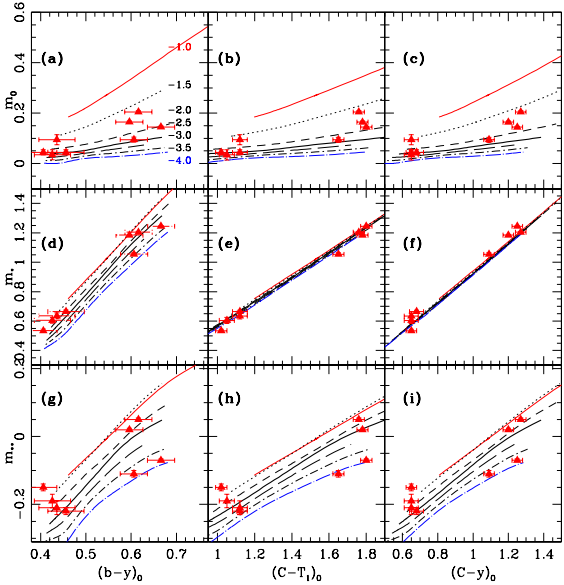}
\caption{Colour-colour plots using a mixture of Str\"{o}mgren and Washington filters to show the [Fe/H]-sensitivity. In all plots, the red triangles are the 8 RGB stars from proper-motion studies \citep{mar07}, and we show the Dartmouth models in the range, $-1.0>[Fe/H]>-4.0$. We define: $m_*=(C-T_1)_0-(T_1-T_2)_0$ and $m_{**}=(C-b)_0-(b-y)_0$, with dereddened-$m_1$ written as $m_0=(v-b)_0-(b-y)_0$.
 \bf (a) \rm $m_0 \; vs.\;  (b-y)_0$.
 \bf (b) \rm $m_0 \; vs.\;  (C-T_1)_0$.
 \bf (c) \rm $m_0 \; vs.\;  (C-y)_0$.
\bf (d) \rm $m_* \; vs.\;  (b-y)_0$.
 \bf (e) \rm $m_* \; vs.\;  (C-T_1)_0$.
 \bf (f) \rm $m_* \; vs.\;  (C-y)_0$.
  \bf (g) \rm $m_{**} \; vs.\;  (b-y)_0$.
 \bf (h) \rm $m_{**} \; vs.\;  (C-T_1)_0$.
 \bf (i) \rm $m_{**}\; vs.\;  (C-y)_0$.
}
\end{figure}

\subsection{SPECTRAL ENERGY DISTRIBUTIONS OF INDIVIDUAL STARS}

When crossing the boundaries between photometric systems, we found it instructive to construct SEDs for our sample, as a independent and external check on our photometry and to make sure all
the magnitudes were on their standard systems \citep{bes05}. This process enabled us to compare the Dartmouth models and the ATLAS9 synthetic stars to the real data, and understand better
the constraints placed by the observational uncertainties and the relationship between each stars' temperature and metallicity. 
Figures 8--15 display SEDs for the 8 RGB proper-motion confirmed members of \boo\  shown with the best-fitting (employing a simple $\chi^2$ fit) ATLAS9 model fluxes, appropriately scaled to pass through the majority of the photometry data points. If there is a wide discrepancy between the fits to the ATLAS9 models in Washington, \str , and SDSS filters, we show 2 fits.
The y-axes are always \it flux density, \rm log($\nu F_\nu$), in ergs/s/cm, and the x-axes are wavelength in microns. We used the conversions from magnitudes to
fluxes from the HST ACS website.\footnote{http://www.stsci.edu/hst/acs/analysis/zeropoints \citep{boh12}} The Washington system and the \str\ data are VEGAmag and the SDSS system is ABmag, and we determined that the 8 RGB stars had the fluxes transformed to the same scale for Washington, \str , and SDSS magnitudes, without any further shifting, except for HWB-22, where the SDSS fluxes are consistently higher. Since the Washington and \str\ data appeared to be consistent with the same $T_{eff}$ for HWB-22, we did not shift any of our zero-point scales. As expected, our photometry uncertainties produce smaller error bars than the SDSS data, and these stars also suffer from large uncertainties in SDSS-u. Where available, we include the \it 2MASS, \rm $JHK$ measurements.

 In all diagrams, the SDSS data is shown as open squares, the \str\ measurements are shown as small filled triangles, Washington magnitudes are the larger filled triangles, and 2MASS data (if available) is shown as open stars. The error bars are the $1\sigma$ uncertainties in the data. If there was a Johnson-V observation, it is shown as an open circle. 

\bf Boo-1137 \rm (Figure 8) is the brightest object in this sample, and the most metal-poor, reported to be $[Fe/H]=-3.7$, by \citet{nor10b}, who  obtained high-resolution spectroscopy of from VLT/UVES.
This star was not originally in our data set because it is 24$^\prime$ from the center of the dSph. \citet{nor10b} pronounced it the most metal-poor giant observed (to-date) in one of the ultra-faint SDSS dSphs, and show that Boo-1137's elemental abundances are similar to those seen in metal-poor Milky Way halo stars.  Given the range in fits from Table 5, we show 
the ATLAS9 model-range which fit the data best, i.e. very metal-poor, $T_{eff}=4750K$, $[Fe/H]=-3.7$, $[\alpha /Fe]=+0.4$, and $\log g=1.4$.

\begin{figure} 
\includegraphics[width=84mm,scale=1.0]{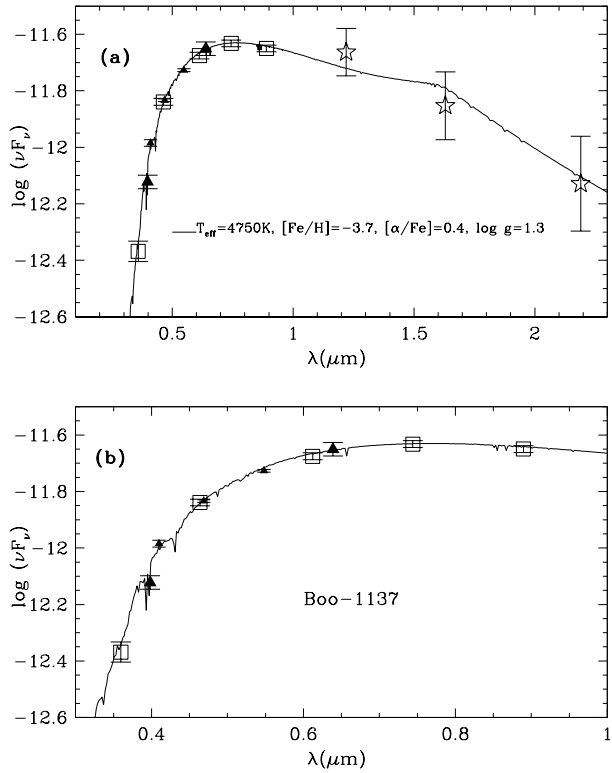}
 \caption{Spectral-energy distribution of Boo-1137 shown with the
best-fitting ATLAS9 (interpolated) model fluxes, appropriately scaled. The SDSS data is
shown as open squares, the Str\"{o}mgren measurements are shown as filled triangles, Washington
magnitudes are the larger filled triangles, and 2MASS data is the open stars. The 
error bars are the $1\sigma$ uncertainties in the data.  \bf (a) \rm SED for the range in wavelength, $0.1\mu m<\lambda <2.3\mu m$. \bf (b) \rm SED for the range in wavelength, $0.3\mu m<\lambda <1.0\mu m$. 
 }
 \end{figure}

\begin{figure} 
\includegraphics[scale=1.0,width=84mm]{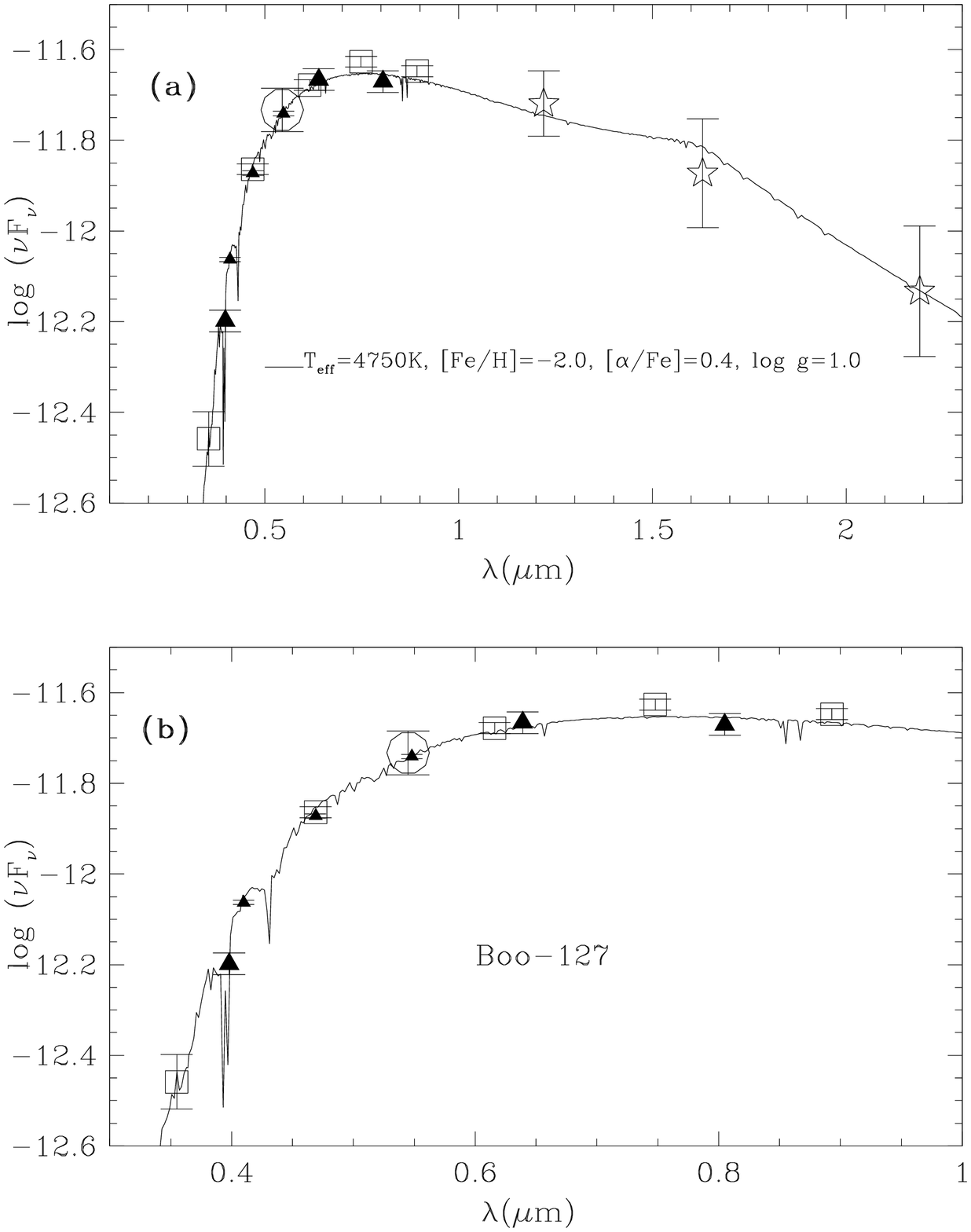}
 \caption{Spectral-energy distribution of Boo-127 shown with the
best-fitting ATLAS9 (interpolated) model fluxes, appropriately scaled. The SDSS data is
shown as open squares, the Str\"{o}mgren measurements are shown as filled triangles, Washington
magnitudes are the larger filled triangles, and 2MASS data is the open stars. The 
error bars are the $1\sigma$ uncertainties in the data. The Johnson-$V$ observation is
is shown as an open circle. \bf (a) \rm SED for the range in wavelength, $0.1\mu m<\lambda <2.3\mu m$. \bf (b) \rm SED for the range in wavelength, $0.3\mu m<\lambda <1.0\mu m$. 
 }
 \end{figure}

\begin{figure} 
\includegraphics[scale=1.0,width=84mm]{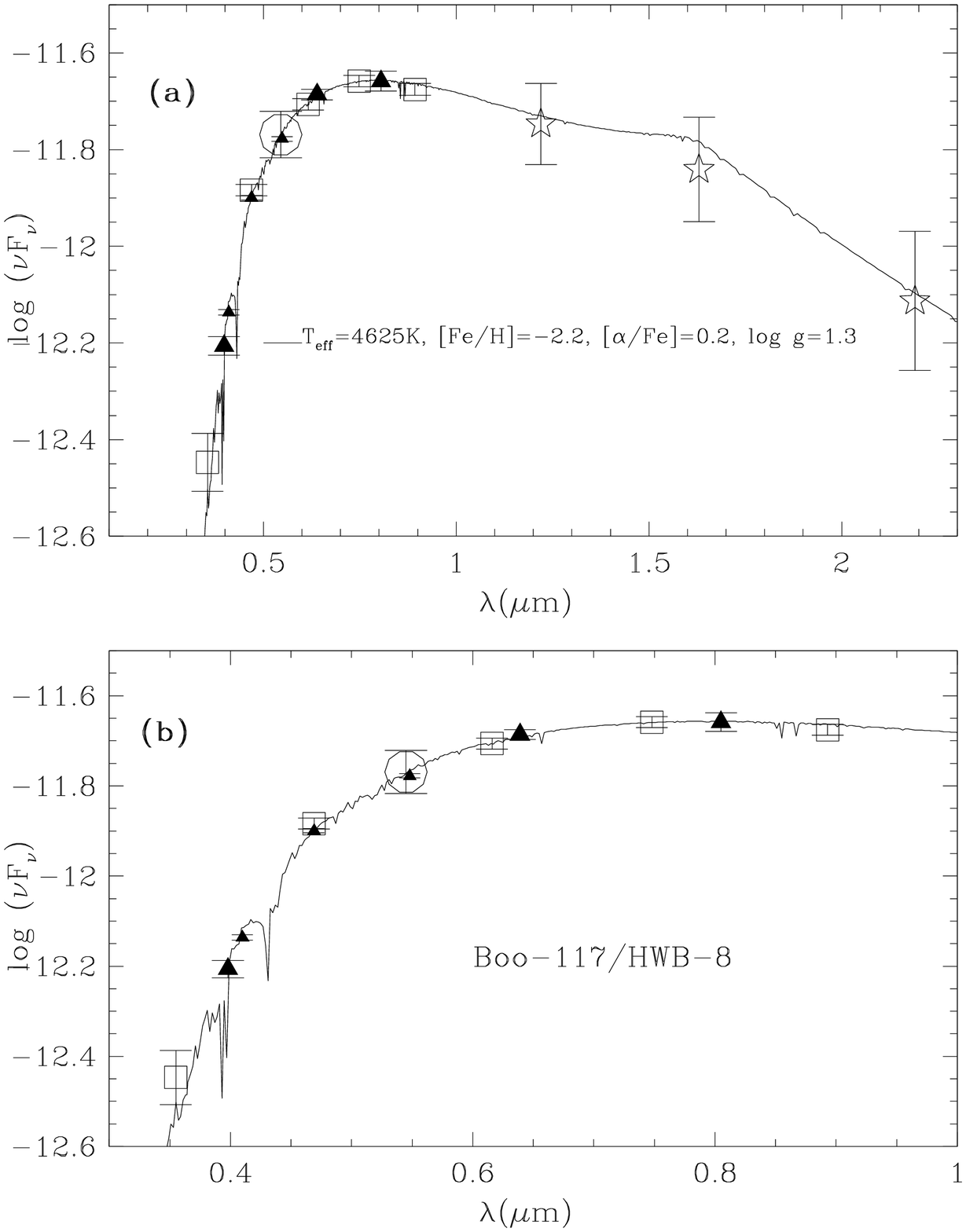}
 \caption{Spectral-energy distribution of Boo-117/HWB-8 shown with the
best-fitting ATLAS9 (interpolated) model fluxes, appropriately scaled. The SDSS data is
shown as open squares, the Str\"{o}mgren measurements are shown as filled triangles, Washington
magnitudes are the larger filled triangles, and 2MASS data is the open stars. The 
error bars are the $1\sigma$ uncertainties in the data. The Johnson-$V$ observation
is shown as an open circle. \bf (a) \rm SED for the range in wavelength, $0.1\mu m<\lambda <2.3\mu m$. \bf (b) \rm SED for the range in wavelength, $0.3\mu m<\lambda <1.0\mu m$. 
 }
 \end{figure}

 \begin{figure} 
\includegraphics[scale=1.0,width=84mm]{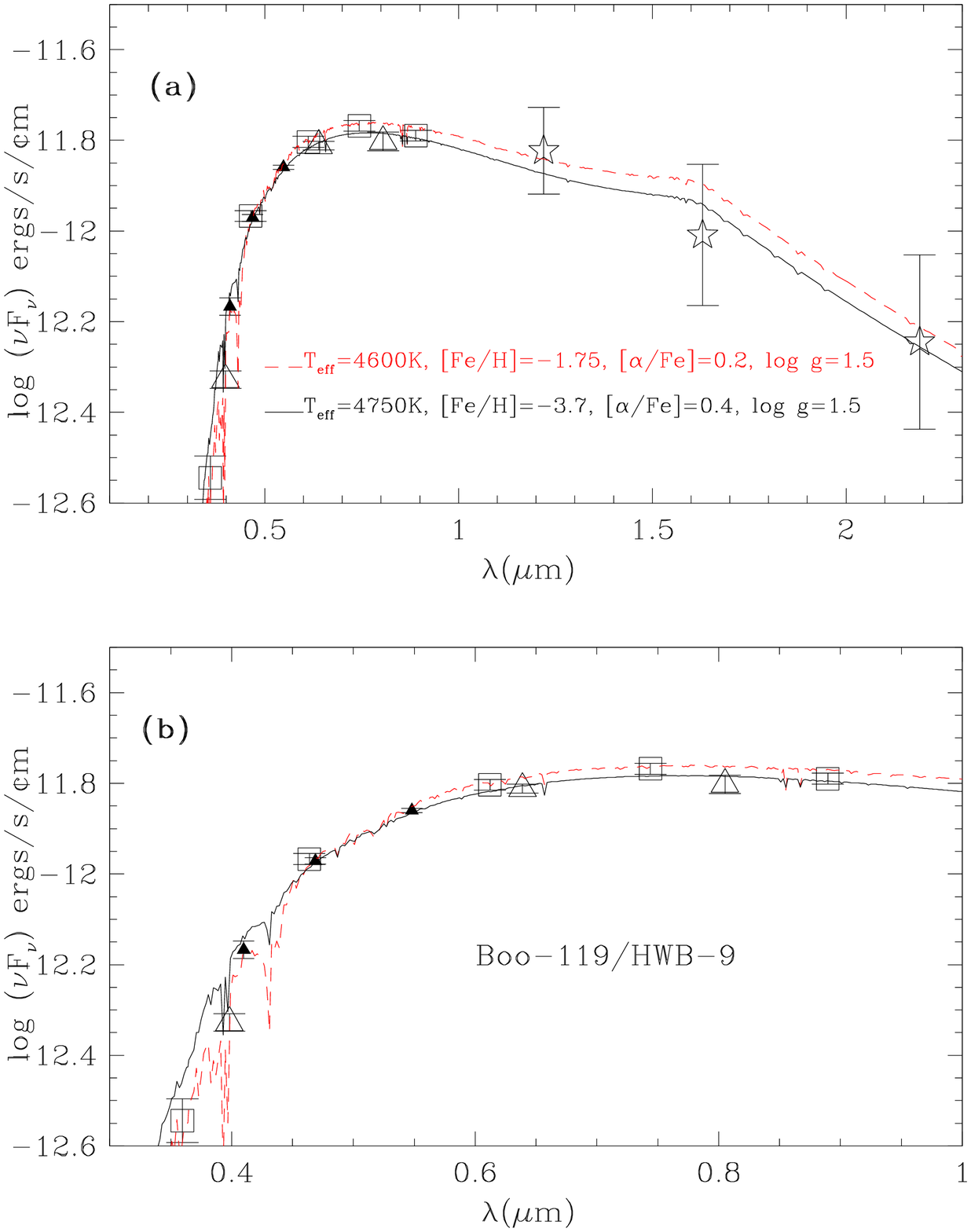}
 \caption{Spectral-energy distribution of Boo-119/HWB-9 shown with the
best-fitting ATLAS9 (interpolated) model fluxes, appropriately scaled. The SDSS data is
shown as open squares, the Str\"{o}mgren measurements are shown as filled triangles, Washington
magnitudes are the larger filled triangles, and 2MASS data is the open stars. The 
error bars are the $1\sigma$ uncertainties in the data.  \bf (a) \rm SED for the range in wavelength, $0.1\mu m<\lambda <2.3\mu m$. \bf (b) \rm SED for the range in wavelength, $0.3\mu m<\lambda <1.0\mu m$. 
 }
 \end{figure}

\begin{figure} 
\includegraphics[scale=1.0,width=84mm]{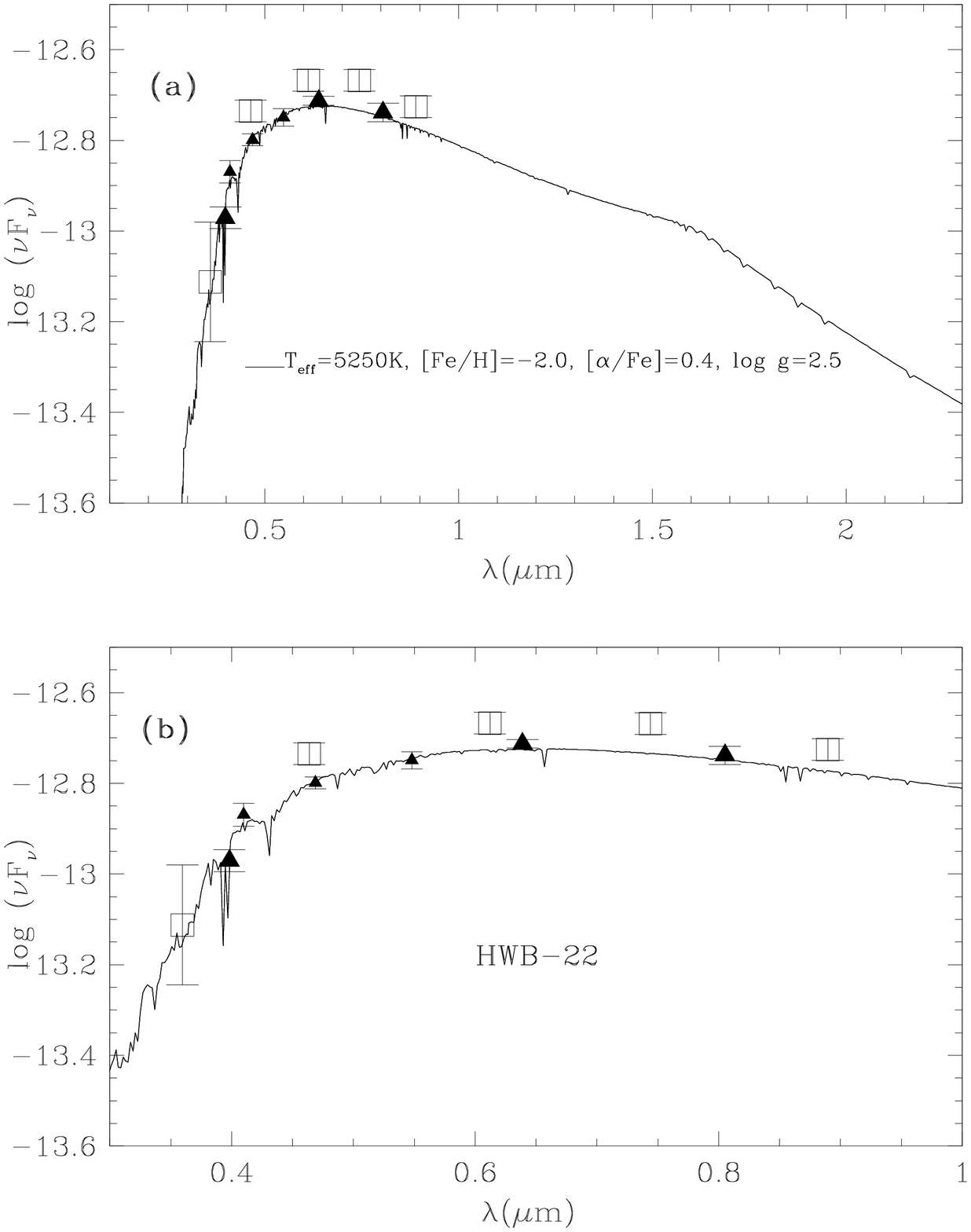}
 \caption{Spectral-energy distribution of HWB-22 shown with the
best-fitting ATLAS9 (interpolated) model fluxes, appropriately scaled. The SDSS data is
shown as open squares, the Str\"{o}mgren measurements are shown as filled triangles, Washington
magnitudes are the larger filled triangles. The 
error bars are the $1\sigma$ uncertainties in the data.  \bf (a) \rm SED for the range in wavelength, $0.1\mu m<\lambda <2.3\mu m$. \bf (b) \rm SED for the range in wavelength, $0.3\mu m<\lambda <1.0\mu m$. 
 }
 \end{figure}

\begin{figure} 
\includegraphics[scale=1.0,width=84mm]{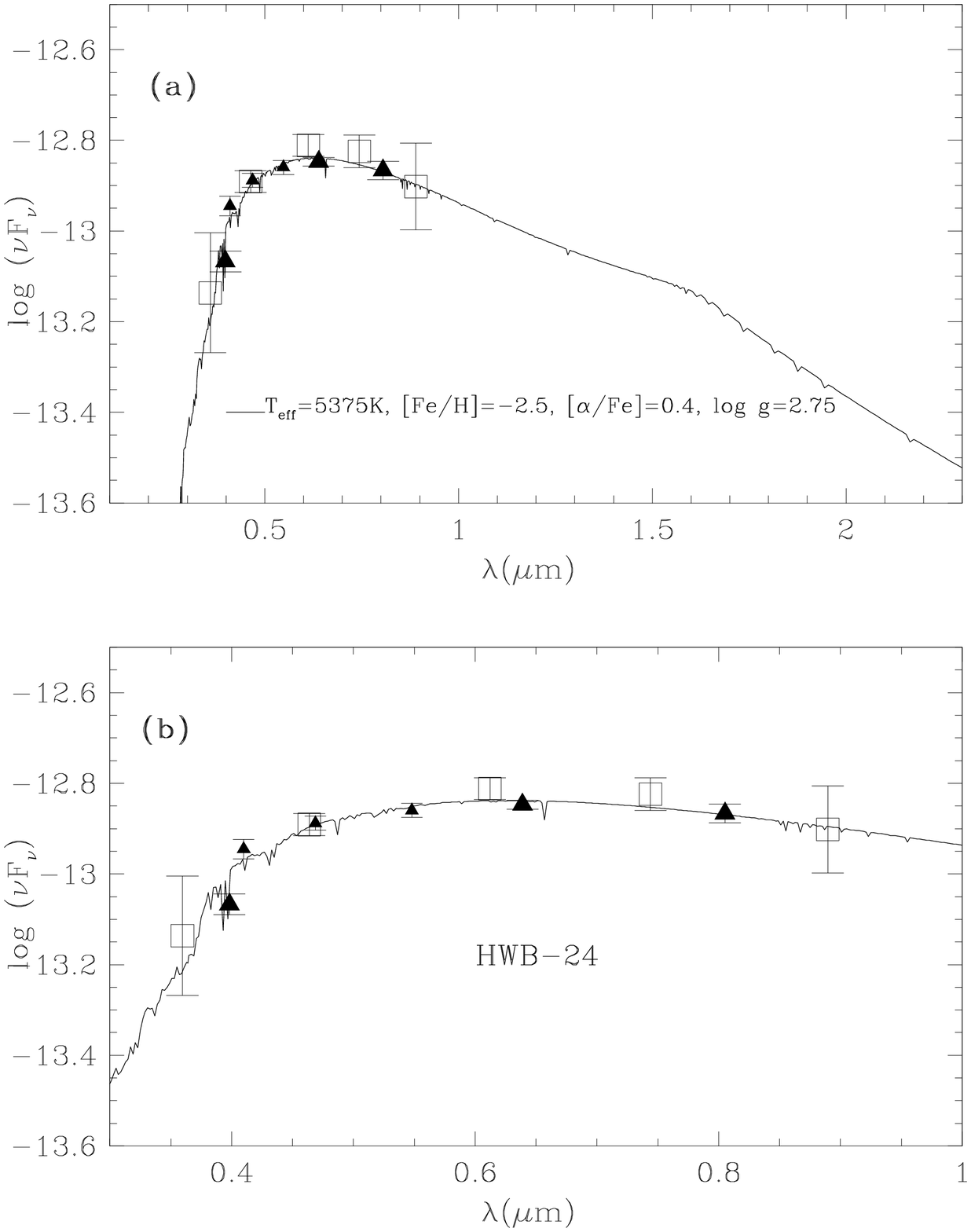}
 \caption{Spectral-energy distribution of HWB-24 shown with the
best-fitting ATLAS9 (interpolated) model fluxes, appropriately scaled. The SDSS data is
shown as open squares, the Str\"{o}mgren measurements are shown as filled triangles, Washington
magnitudes are the larger filled triangles. The 
error bars are the $1\sigma$ uncertainties in the data.  \bf (a) \rm SED for the range in wavelength, $0.1\mu m<\lambda <2.3\mu m$. \bf (b) \rm SED for the range in wavelength, $0.3\mu m<\lambda <1.0\mu m$. 
 }
 \end{figure}

\begin{figure} 
\includegraphics[scale=1.0,width=84mm]{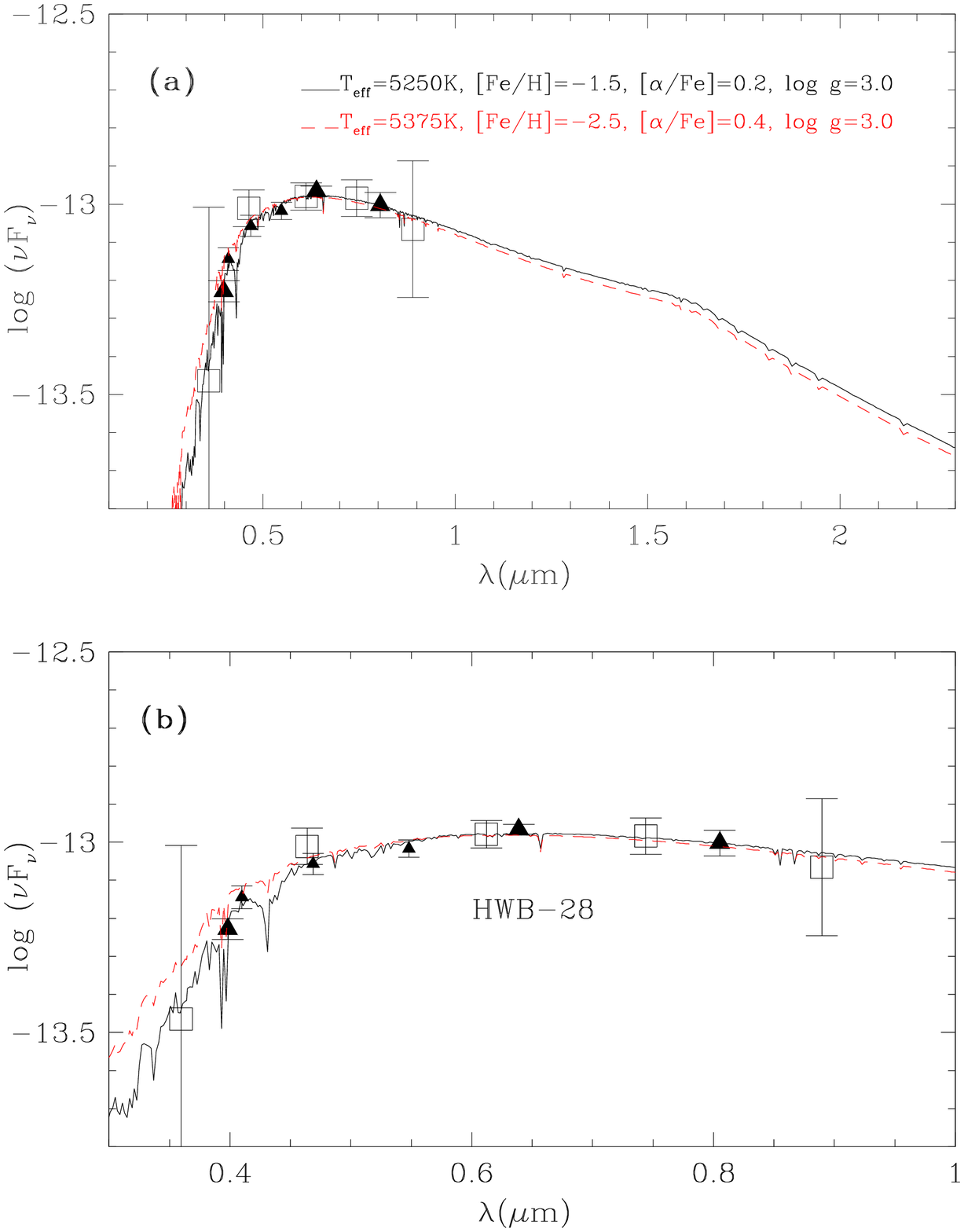}
 \caption{Spectral-energy distribution of HWB-28 shown with the
best-fitting ATLAS9 (interpolated) model fluxes, appropriately scaled. The SDSS data is
shown as open squares, the Str\"{o}mgren measurements are shown as filled triangles, Washington
magnitudes are the larger filled triangles. The 
error bars are the $1\sigma$ uncertainties in the data. \bf (a) \rm SED for the range in wavelength, $0.1\mu m<\lambda <2.3\mu m$. \bf (b) \rm SED for the range in wavelength, $0.3\mu m<\lambda <1.0\mu m$. 
 }
 \end{figure}

\begin{figure} 
\includegraphics[scale=1.0,width=84mm]{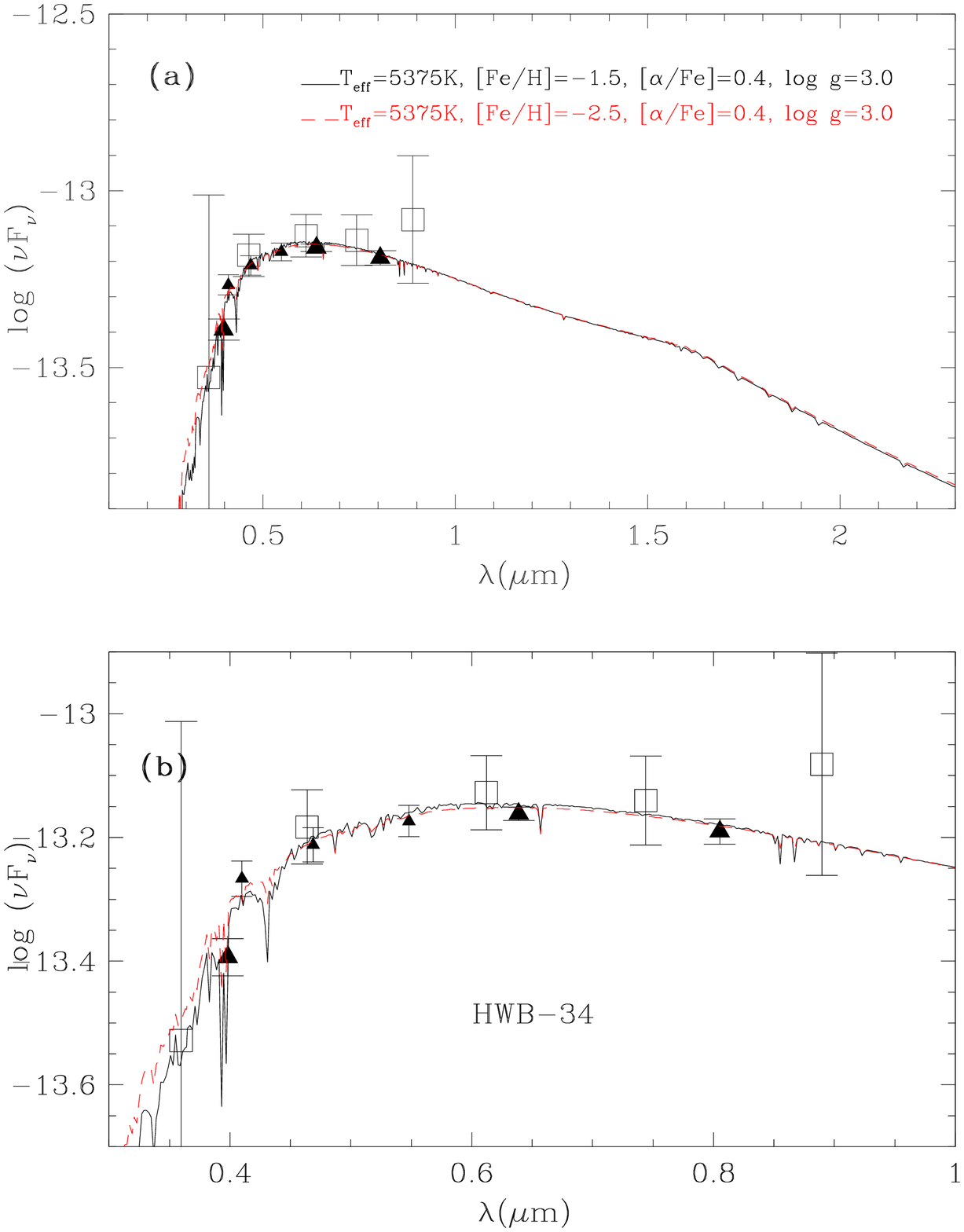}
 \caption{Spectral-energy distribution of HWB-34 shown with the
best-fitting ATLAS9 (interpolated) model fluxes, appropriately scaled. The SDSS data is
shown as open squares, the Str\"{o}mgren measurements are shown as filled triangles, Washington
magnitudes are the larger filled triangles. The 
error bars are the $1\sigma$ uncertainties in the data. \bf (a) \rm SED for the range in wavelength, $0.1\mu m<\lambda <2.3\mu m$. \bf (b) \rm SED for the range in wavelength, $0.3\mu m<\lambda <1.0\mu m$. 
 }
 \end{figure}

\bf Boo-127 \rm  (Figure 9) was found to have an unusual [Mg/Ca] abundance ratio, similar to the Hercules dSph \citep{fel09}. \citet{fel09} found $[Fe/H]=-1.98$, while \citet{nor08} found $[Fe/H]=-1.49$. From the spectroscopy in Table 5 and isochrone fits in Table 9, we find that  $[Fe/H]=-2.0$ across all the filter-systems and that the best ATLAS9 fits are models, with $T_{eff}=4750K$, $[Fe/H]=-2.0$, $[\alpha /Fe]=+0.4$, and $\log g=1.3$.  
However, \citet{nor10a} later obtained UVES/FLAMES data for Boo-127, where they reported, $[Fe/H] = -2.08$; with a mean error of about 0.27 dex. \citet{gil13b} concurs, so we can conclude that the spectroscopic $[Fe/H]$-values are now in agreement. 

\bf Boo-117/HWB-8 \rm  (Figure 10) has the most independent observations.  \citet{nor08} measured $[Fe/H]=-1.72$, using the calibration from the strength of the Ca II K line, and found $[Ca/Fe]\sim 0.3$. This value was more metal-rich than the $[Fe/H]=-2.2$  noted by \citet{mar07}, and is at least 0.4 dex more metal-rich than any of our other photometric estimates. \citet{fel09} reported a value of $[Fe/H]=-2.24$  for Boo-117 (HIRES; using the Fe lines, amongst others), that is more consistent with our \str\ data. As discussed in \citet{fel09}, even different spectroscopic surveys can vary from $0.3$ to $0.5$ dex, so we should not be surprised by this.  We show the ATLAS9  models/interpolations, with $T_{eff}=4625K$, $[Fe/H]=-2.25$ \&  $[\alpha /Fe]=+0.2$ \&  $\log g=1.25$. We note that the bluer photometry points are more consistent with
a slightly more metal-rich spectrum, but that all the fits are well-within the uncertainties stated in the spectroscopic studies. Also, \citet{lai11} derive $[Fe/H]=-2.3$ and $[C/Fe]=-0.5$, from their low resolution spectra, SDSS and other filters  with the n-SSPP code.

\bf Boo-119/HWB-9 \rm  (Figure 11) is a star we expected to have inconsistent photometric estimates, since \citet[][Boo-119]{mar07} derived
$[Fe/H]=-2.7$, and the isochrone fits to each filter set range into the more metal-rich regime: $[Fe/H]=-2.0$ for the SDSS filters, The Washington filters give $[Fe/H]=-1.7$, and we find $[Fe/H]=-1.5$ from the \str\ isochrones. We expected this discrepancy to be a characteristic of a CN- or carbon-rich star, and \citet{lai11} report $[Fe/H]=-3.8$ and $[C/Fe]=+2.20$.  The best-fit ATLAS9 models indicate an interpolation in the
range $T_{eff}=4625K-4750K$, $[Fe/H]=-1.8$ to $-2.5$, $[\alpha /Fe]=+0.2$ to $+0.4$, and $\log g=1.5$. The model with $T_{eff}=4750K$ and $[Fe/H]=-2.5$, has the smallest residuals, and is consistent with the CaT-measurement from \citet{mar07}. We show the range of
models with $T_{eff}=4600-4750K$, $[Fe/H]=-1.75$ to $-3.7$, $[\alpha /Fe]=+0.2$ to $+0.4$, and $\log g=1.5$.

\bf HWB-22 \rm  (Figure 12) may be variable. There is no clear offset in the photometry points between the different filter
systems for any of the other RGB stars, but it does seem be the case here. \citet{mar07} give $[Fe/H]=-2.2$, \citet{lai11} give $[Fe/H]=-2.9$ and $[C/Fe] < 0.0$, and our isochrone fits give $[Fe/H]=-2.0$ to $-3.0$. The ATLAS9 models indicate the best
fit is: $T_{eff}=5250K$, $[Fe/H]=-2.0$, $[\alpha /Fe]=+0.4$,  and $\log g=2.5$. The difference in photometry could be caused by a temperature difference, but that the metallicity  and surface gravity measurements are all consistent with each other. We chose to show the fit
to the \str\ and Washington colours.

\bf HWB-24 \rm  (Figure 13)  The \citetalias{gei99} calibration gives $-3.0\pm 1.0$ dex, and adds no further information. The \citetalias{gei99} standard
RGB's (and the \citealt{dot08} models in Washington filters) become insensitive to very low metallicities amongst the brighter RGB stars much more rapidly than the \str\ calibrations. For Washington data on low-metallicity 
systems, it is better to use the CMDs than the colour-colour plots, since the models separate better
in luminosity and $(C-T_1)$ than they do in $(T_1-T_2)$. At this level on the RGB, the \str\ calibration starts to lose sensitivity.

\bf HWB-28 \rm (Figure 14) fits the SDSS isochrones and the \citet{lai11} models both with$[Fe/H]\geq -2.5$, but the \str -fit 
gives $[Fe/H]=-1.7$. The fit from the Washington filters gives $[Fe/H]=-1.5$, closer to the \citet{mar07} estimate $[Fe/H]=-1.5$ from the CaT-method.
From the ATLAS9 models, we can see the  why the SDSS filters can indicate a much more metal-poor model than the \str\ filters
and the C-band. Note that the g-band error bars encompass both model SEDs, $T_{eff}=5250K-5375K$, $[Fe/H]=-1.5$ and $-2.5$, $[\alpha /Fe]=+0.2$ and $+0.4$,  and $\log g=3.0$. This temperature range makes the dashed line fit the SDSS photometry with smaller residuals than the more metal-rich case.

\bf HWB-34 \rm (Figure 15) has a consistent ATLAS9 model temperature, with $T_{eff}=5375K$, and we show $[Fe/H]=-1.3$ and $-2.5$, $[\alpha /Fe]=+0.2$ and $+0.4$ ,  and $\log g=3.0$. \citet{lai11} fit $[Fe/H]=-2.4$, \citet{mar07} find  $[Fe/H]=-1.3$. The \str\ system isochrones yield $[Fe/H]=-2.0$ and the $CT_1T_2$-photometry gives $[Fe/H]=-1.7$. This star is at the exactly the right (or in this case, wrong) part of the temperature/metallicity distribution to make it very difficult to discern where this star falls within $[Fe/H]=-1.7\pm 0.5$ dex, and the deciding factor would be the $C$- or $v$-magnitude.

\subsection{COMPARISON WITH SIMPLE MODELS}

 For the 8 \boo\  radial-velocity-confirmed members (Table 5), the photometric metallicities that were most discrepant were for the fainter objects on the lower RGB, as we expected. However, Boo-119/HWB-9's \rm  (Figure 11)
discrepancies are likely due to carbon-enhancement. We now compare the spectroscopic metallicities to those found from different filter sets and hybrid methods.  Table 5 shows a clear offset from the [Fe/H]-values from the spectroscopy-alone studies \citep{gil13b,nor10b,fel09,mar07,hug08}, but that shift
to lower $[Fe/H]$ was discussed in \S 2, and is partly to do with scaling changes over time and calibration of CaT-measurements.
 Apparently, no photometric system can cope with Boo-119's carbon-iron abundance ratio, $[C/Fe]=+2.20$, except for \citeauthor{lai11}'s \citeyearpar{lai11} modified n-SSPP method, which skews all [Fe/H]-estimates lower than any scale based on CaT alone (\citet{kop11} notes that their scale is not calibrated below $[Fe/H]=-2.5$)

\citet{fel09} discuss their spectroscopy of \boo , in light of work by \citet{koc08} on the Hercules dSph.
In these dark-matter dominated dSphs, the very low density of baryons should mean that the chemical enrichment history is dominated 
by only a ``few supernovae", which means that the element abundances of the stellar population might show star-to-star chemical inhomogeneities, tracing
individual SNe II events \citep{gil13a,gil13b}. In studying the Hercules system, \citet{koc08} estimate that only 10 SN were needed
to give the ``atypical abundance ratios" (and more should have contributed).  \boo\ is less massive than Hercules, but \citet{fel09} only report one star,   Boo-127, (and possibly Boo-094, also not observed here), has  unusual variations in Mg and Ca. Expecting more individuality, \citet{fel09} conclude that \boo\ is surprisingly well-mixed. \citet{nor09} beg to differ, and they take issue with \citeauthor{fel09}'s \citeyearpar{fel09} $[Mg/Ca]$ value as being too high for one of their seven \boo\ stars, the aforementioned, Boo-127. We agree with Norris's group, and also \citet{gil13b}.

\citet{nor09} obtained high-resolution spectroscopy of  Boo-1137 from VLT/UVES  (not
included in \citetalias{hug08}'s data set because it is $24^\prime$ from the center of the dSph), pronouncing it 
the most metal-poor giant observed (to-date) in one of the ultra-faint SDSS dSphs, with $[Fe/H]=-3.7$. \citet{nor09} find that Boo-1137's elemental abundances are similar to those seen in metal-poor MWG halo stars. They also discuss \boo 's SF history, surmising that it most likely underwent
an early, short period of star formation, cleared by SN II which enriched it's
interstellar medium inhomogeneously. In Boo-1137, \citet{nor09} find $\alpha$-enhancement
(compared to Fe) of $\Delta[\alpha/Fe]\sim 0.2$, ``relative to the mean of the metal-poor halo stars."
Even though the spectroscopy studies we have mentioned have tried to cover as much of the \boo\  dSph as possible (and Boo-1137 
is almost two half-light radii from the center), all groups have collected  medium resolution spectra
of less than 30 members, with high-resolution spectra of only a handful of stars. As \citet{tol09} and \citet{gil13a} say, SNe Ia start to contribute to the chemical composition $10^8-10^9$
years after the first stars form. There is a wide range in  $[Fe/H]$, but not in the $\alpha$-elements.
So, if the massive stars in \boo\ cleared the interstellar medium quickly, star formation lasted less
than 1~Gyr. Can we detect an age spread that small?

\begin{table*}
\begin{minipage}{156mm}
\caption{Metallicities of Possible \boo\ Stars}
\begin{tabular}{@{}lccccccccccl}
\hline 
{ID}& 
{$[Fe/H]_{Hil}^1$}& {$[Fe/H]_{m_1}^2$}&  {$[Fe/H]_{[m]}^3$}&    {$[Fe/H]_{GS}^4$}& \multicolumn{4}{c}{$[Fe/H]_{spec}^5$}& {$[\alpha/Fe]_{Gil}^6$}&
{$[Fe/H]_{Lai}^7$}& {$[C/Fe]_{Lai}^7$}\\
\hline
Boo-1137& $-2.1(0.3)$& -2.4& -2.3& $-2.2(0.3)$& -3.7&$-$&  -3.66& y& 0.44& $-$& $-$ \\
Boo-127& $-2.0(0.3)$& -2.2& -2.2& $-2.0(0.3)$& -1.49& $-$& -2.01& y& 0.18& $-$& $-$ \\
Boo-117/HWB-8$^8$& $-1.6(0.3)$& -1.9& -1.8&  $-2.0(0.3)$& -1.72& -2.2& -2.18&  y& 0.18& -2.3& -0.50 \\
Boo-119/HWB-9& $-1.5(0.3)$& -1.7& -1.7&  $-2.0(0.3)$& $-$& -2.7&  -3.33& y& 0.77& -3.8& +2.20    \\
HWB-16& $-1.8(0.6)$& -2.2& -2.2&  $-2.6(0.5)$& $-$& $-$& $-$& u& $-$& $-$& $-$       \\
HWB-19& $-0.5(1.0)$& -0.5& -0.6& $-3.3(0.8)$& $-$& $-$& $-$& n& $-$& $-$& $-$ \\
HWB-22& $-1.8(0.6)$& -2.3& -2.3&  $-2.9(0.7)$& $-$& -2.0& $-$& y& $-$& -2.9& $<0.00$ \\
HWB-24& $-1.4(0.7)$& -2.0& -2.0&  $-3.0(0.7)$& $-$& -1.9& $-$& y& $-$&  -2.3& +0.40  \\
HWB-28& $-1.2(0.6)$& -1.6& -1.6&  $-2.9(0.6)$& $-$& -1.5& $-$& y& $-$& -2.5& +0.39   \\
HWB-29& $-1.4(0.9)$& -1.8& -1.7&  $-2.8(0.6)$& $-$& $-$& $-$& u& $-$& $-$& $-$\\
HWB-31& $-2.3(0.9)$& -3.8& -3.6&  $-3.1(0.8)$&      $-$&  $-$& $-$& u& $-$& $-$& $-$ \\
HWB-34& $-1.7(0.9)$& -2.3& -2.3&  $-3.0(0.7)$& $-$& -1.3& $-$& y& $-$&-2.4& +0.34   \\
HWB-40& $-1.6(0.9)$& -2.1& -2.1&  $-2.9(0.7)$& $-$& $-$& $-$& u& $-$& $-$& $-$       \\
\hline
\end{tabular} \\
(1) From photometry: \str -$[Fe/H]$ calibration from \citet{hil00}.\\
(2) From photometry: \str -$[Fe/H]$:-- $m_1$ calibrations from \citet{cal07}, uncertainties are $\sim 60 \%$ greater than the \citet{hil00} calibration.\\
(3) From photometry: \str -$[Fe/H]$:--  [m] calibrations from \citet{cal07}.\\
(4) From photometry: averaging \citetalias{gei99}'s standard RGB-calibrations, since there is a spread in metallicity.\\
(5) From spectroscopy, respectively: \citet{nor08,mar07,gil13b}. Proper motion membership designated: yes=y; no=n; unknown=u.\\
(6) \citet{gil13b}\\
(7) From photometry and spectroscopy \citet{lai11} n-SSPP method.\\
(8) Boo-117 appears in \citet{mar07,nor08,fel09}. \citet{gil13b} prefers $[Fe/H]\sim -2.2.$
 \end{minipage}
\end{table*}

\begin{table*}
\begin{minipage}{110mm}
\caption{Fitted Parameters for \boo\ Stars from \citet{hug11b}}
\begin{tabular}{@{}lccccc}
\hline 
ID & $[Fe/H]_{Spec}^1$ & $[Fe/H]_{phot}^2$  & $T_{eff}(K)^3$& $\log g^4$& Age (Gyr)$^5$\\
\hline
Boo-117/HWB-8 & -2.25 & -2.4& 4700 & 1.4 & 11\\
Boo-119/HWB-9 & -2.7& -2.7& 4750 & 1.5& 12 \\
HWB-22& -2.2 &-2.2 & 5250 & 2.6 & 12 \\
HWB-24 & -1.9& -1.8& 5300 & 2.7& 12 \\
HWB-28 & -1.5& -1.2 & 5350 & 2.9& 11\\
HWB-34& -1.3& -1.1& 5400& 3.1& 12 \\
\hline
\end{tabular} \\
(1) From \citet{mar07}.\\
(2) Means found from fitting Washington and \str\ photometry separately to the $\alpha$-enhanced
Dartmouth isochrones, then taking weighted means, dependent on error bars. Uncertainties
are then $\pm 0.25$ dex.\\
(3) Weighted mean from Dartmouth isochrones, uncertainties are $\pm 50K$.\\
(4) Weighted mean from Dartmouth isochrones, uncertainties are $\pm 0.05$ dex.\\
(5) Weighted mean from Dartmouth isochrones, uncertainties are $\pm1.0$ Gyr.\\
 \end{minipage}
\end{table*}

\begin{table*} 
\begin{minipage}{80mm}
\caption{Model Fits for $-1.8>[Fe/H]>-2.5$ and Constant Age of 11.5 Gyr, $[\alpha/Fe]=0.0$}
\begin{tabular}{@{}lccc}
\hline 
$\Delta$colour/Index& MSTO &  RGB   & rRGB\\ (mag.)& $M_{T_1}\sim 3$& $M_{T_1}\le -1.5$&
$2>M_{T_1}>1$\\
\hline
$(C-T_1)$  &0.267 &0.557 &0.247\\
$(T_1-T_2)$  &0.080 &0.106 &0.047\\
$(b-y)$  &0.073 &0.157 &0.057\\
$m_1$  &0.024 &0.115 &0.045\\
$[m]$  &0.014 &0.162 &0.054\\
$m_*$  &0.187 &0.451 &0.199\\
$m_{**}$   &0.051 &0.133 &0.092\\
$(C-y)$  &0.196 &0.447 &0.202\\
$(u-g)$  &0.039 &0.519 &0.192\\
$(g-r)$  &0.139 &0.221 &0.099\\
$(r-i)$  &0.071 &0.091 &0.064\\
$(B-V)^\dagger$ & 0.203& 0.341& 0.193\\
$(V-R)^\dagger$ & 0.122& 0.201& 0.116\\
$(V-I)^\dagger$ & 0.252& 0.401& 0.242\\
\hline
\end{tabular}
\\Non-$\alpha$-enhanced models generated from the Dartmouth isochrones \citep{dot08}. $^\dagger$From Table 4 of \citet{gei96}, converting $(C-T_1)$ to BVRI colours.
 \end{minipage}
\end{table*}

\it What kind of metallicity or age spread would we expect to see in \boo ? \rm
Age estimates cannot usually be made for individual RGB stars, but the \citetalias{gei99} standard
giant branches do have an age-metallicity degeneracy, compared to the Dartmouth isochrones. Is the separation of RGB stars with \boo 's range in chemical composition, large enough to differentiate between RGB-isochrones
in any colour? Our normal practice is to use the RGB to find the metallicity and the MSTO to find ages, where the isochrones
separate.  \citet{bro12} and \citet{kir12} studied the metallicity distributions and 
star formation histories (SFHs) of several dSphs. Isochrone fits to these populations, as noted by Kirby,
have to be performed at the upper-RGB, the SGB and the MSTO, in order to break the age-metallicity degeneracy. In many dSphs, we do not have sufficient upper-RGB stars to make this fit \citep{wil10}.

\begin{table*} 
\begin{minipage}{80mm}
\caption{colour Change/Gyr for Model Fits for Constant $[Fe/H]=-1.8$, $[\alpha/Fe]=0.0$, and Age Range of 10-14 Gyr}
\begin{tabular}{@{}lccc}
\hline 
$\Delta$colour/Index& MSTO &  RGB   & rRGB\\ (mag.)& $M_{T_1}\sim 3$& $M_{T_1}\le -1.5$&
$2>M_{T_1}>1$\\
\hline
$(C-T_1)$&      0.129&  0.417&  0.016\\
$(T_1-T_2)$&    0.040&  0.084&  0.004\\
$(b-y)$ &       0.037&  0.109&  0.004\\
$m_1$   &       0.010&  0.135&  0.000\\
$[m]$   &       0.004&  0.168&  0.001\\
$m_*$   &       0.090&  0.333&  0.012\\
$m_{**}$&       0.027&  0.111&  0.003\\
$(C-y)$ &       0.095&  0.329&  0.012\\
$(u-g)$ &       0.008&  0.413&  0.014\\
$(g-r)$ &       0.072&  0.170&  0.007\\
$(r-i)$ &       0.008&  0.076&  0.003\\
$(B-V)^\dagger$ & 0.137& 0.274& 0.084\\
$(V-R)^\dagger$ & 0.084& 0.163& 0.053\\
$(V-I)^\dagger$ & 0.181& 0.329& 0.123\\
\hline
\end{tabular}
\\Non-$\alpha$-enhanced models generated from the Dartmouth isochrones \citep{dot08}. $^\dagger$From Table 4 of \citet{gei96}, converting $(C-T_1)$ to BVRI colours.
\end{minipage}
\end{table*}

To improve the resolution of our grid of Dartmouth isochrones, we ran a simple closed-box chemical evolution mode. We used the Dartmouth (solar-scaled)  isochrones to produce  populations of stars with a similar physical size, mass, and metallicity spread
to \boo , but with ages of 10--14~Gyr over the approximate metallicity range  $-1.0>{Fe/H}>-4.0$. We obtained over 50,000 artificial stars (binary fraction of 0.5) with \str , Washington and
SDSS magnitudes. We compared model stars with our
objects having $vbyCT_1T_2$ measurements. 

From these model runs, we chose 3 simple cases. Firstly, there is a single-age (11.5~Gyr) population (1216 stars), with a conservative spread in metallicity from $[Fe/H]=-3.5$ up to 
$-1.5$. We call this Case I.
The second population (Case II: 9133) was designed to be enriched \it in situ, \rm with $-4.0 < [Fe/H] <-1.0$. The metallicity distribution  terminated after star formation lasting at least 4~Gyr. We added Case III, an alternative long-duration population with a few ``starbursts,'' to determine if we could tell it apart from Case II. In Case III, there are
1374 stars born 12~Gyr ago, with $-3.5<[Fe/H]< -2.5$, 6375 stars with $-2.5< [Fe/H]< -1.75$, which are 0.5~Gyr younger, and 1363 stars  with $-1.75< [Fe/H]< -1.0$, with SF extending up to 10~Gyr ago, and then terminating. Using the Kolmogorov-Smirnov Test for Two Populations, where the null-hypothesis ($H_0$) is that the populations are the same, we can tell the difference between the 
age spreads in Cases I, II and III, but not the metallicity ranges.
  
In order to make the models more realistic, we use the photometric uncertainties from Figure 3. From tests, we found that we required at least 2\%  photometry at the MSTO to determine if there are age spreads present, since the isochrones  exhibit a change of $\sim 0.06$ mag. in $(C - T_1)$ for each~Gyr in age. There would have to be much deeper Str\"{o}mgren photometry for any age spread less than 1.5~Gyr to be distinguished. In the Washington system alone,
with $[Fe/H]\sim -2.0$, 2 per cent  photometric uncertainty at the MSTO would reveal an age spread of $> 1$~Gyr.

We selected models with $\log g$ of 1.0-2.0 for the RGB stars, 2.0-3.0 for the rising-RGB (rRGB), below the level of the HB, and $\log g\sim 4.0$ for the MSTO. We examined the colours in the \str , Washington and SDSS systems, at fixed age and fixed-[Fe/H]. Our goal was to quantify the filter systems sensitivity and break the age-metallicity degeneracy as cleanly as possible, avoiding the
upper-RGB.
The results of the artificial star experiments  are given in Tables 7 and 8, and Figure 16.
Figure 16a shows colours and indices for the stars in all the models with an age of 11.5~Gyr, and examines a range of $-2.5< [Fe/H]< -1.8$. The RGB stars' range is shown in orange, the rRGB in gold, and the MSTO in violet. Figure 7b takes stars at a constant $[Fe/H]=-1.8$. We know that the real stars are $\alpha$-enhanced at the +0.2 to +0.4 dex level, \citetext{\citealt{nor10b} and \citealt{gil13b}; see Table 5} which would make a metal-poor star look more metal rich in the broad-band colors. In Figure 16b, we let the model stars have a fixed metallicity value of $[Fe/H]=-1.8$ (more like a real sample of $[Fe/H]\ge -2.0$) and let the age vary from 10-14~Gyr. 

Even though these models cannot be considered well-calibrated below $[Fe/H]=-2.5$, we show the lower end of the $[Fe/H]$-range, from $-4.0<[Fe/H]<-3.0$, at a constant age of 11.5~Gyr, but we 
hatch the bar-chart to recognize this factor. If we were fortunate to collect so many $[Fe/H]$-values of enough real stars in dSphs at this low range in $[Fe/H]$, the stars may be $\alpha$-enhanced and may not appear so metal-poor broadband colors (CN- or C-enhanced). However, this
is the parameter-space where the SDSS-colors recover their usefulness. 
For our sample, that the SDSS-colors are very uncertain here, as the $u-$ and $g-$magnitudes have large uncertainties (see Table 4) returned are too low for the MSTO-stars, at the sensitive TO-value of $\log g$. If we had detections down to the MSTO stars, they would have a spread in $(u-g)\ge 3.0$. Here, the SDSS filters lack of sensitivity to $\alpha$-elements lets us recover a star's metal-poor nature, even if it was an extreme example, e.g. Boo-119. However, for this color to be effective, the stars have to be \it detected \rm at $SDSS-u$, which takes more than twice as long as the $C$-band (see Table 2).
For stars just above the MSTO to the upper-RGB, $(C-T_1)$ is most effective for all evolutionary stages in both age and metallicity. The $(C-T_1)$ calibration loses is sensitivity below $[Fe/H]\sim -2.2$ on this scale, which is why we have to supplement it with $(b-y)$. The filter combination $CT_1by$, or even $Cby$, maintains its sensitivity to metallicity below $[Fe/H]\sim -3.0$, as can be seen in Figures 7g--i. 
 Also note that the rRGB colours are hardly sensitive to age at all, and can be used to break the age-metallicity degeneracy. We want to avoid using Sloan-u or \str -u, and the $v$-band, as both consume observing time (see Table 2).  However, many searches for closer and brighter extremely metal-poor stars \citep{spi13}, this color could be vital, but even they use $(g-z)$ to avoid SDSS-u. 
 
 From the models, we find that the $(C-T_1)$-colour is best for our sample above the MSTO;  the $(C-y)$-colour is shown to be useful (in agreement with \citealt{ros14}), $m_*$ is better than $m_{**}$, and all these are better than the SDSS colours. However, constructing indices involves 3 or 4 filters, which increases the uncertainty in an parameters derived from a colour-colour or color-index calibration. The method of deriving age and $[Fe/H]$ will the smallest uncertainty is to use a color sensitive to temperature and one sensitive to chemical composition, but we need to combine the Washington and \str\ filters.

We considered that \boo\  might resemble $\omega$ Cen, the core of a captured dE, with extended enrichment and a long period of star formation \citep{gra11,hug04}.
Due to its lack of H{\scriptsize I} or recent star formation (SF) \citep{bai07}, we might expect \boo\ to have a simple history, similar to most globular clusters \citep{mil12}, with a short burst of star formation, terminated by supernovae (SNe)  II events, or tidal stripping \citep{fel08} removing the ISM. This scenario should give rise to almost a single-age fit to the CMD, and a very narrow ($\Delta [Fe/H] <0.5$) dex metallicity range. \citet{mar08} discussed the model of \citet{mar06} which describes the chemical evolution of dSphs (particularly Draco, but it can be generalized). \citet{mar08} argue that the SNe II ejecta from the star formation event in a Draco-like dSph (actually about  ``50 instantaneous bursts'' over 60 Myr) should be retained, due to the extra mass from the dark matter
halo, and because ``efficient radiative losses" keep the enriched ISM flowing back to the central regions of the dSph, mixing the gas. \citet{coh09} obtained high-resolution
spectra of 8 stars in Draco, adding to 14 they had observed previously, which had a metallicity range of $-1.5>[Fe/H]>-3.0$ dex, similar to \boo ;
their age-metallicity relationship indicated that SF in Draco lasted about 5~Gyrs. An age spread even half this large should have been obvious for
\boo , even with the uncertainties in Figure 3.
The modeling  of Draco's enrichment
history \citep{mar08}, showed a star formation process lasting about 2~Gyr, with most of the initial Fe-enrichment coming from SNe II,
with chemical inhomogeneities appearing later, caused by SNe Ia ``pockets" of material, with higher $[Fe/H]$ and lower $[\alpha/Fe]$, which
take 2--3~Gyr to mix in with the rest of the gas, the dynamics of which is driven by the SNe II effects. The discussion of possible star formation histories for this system is also illustrated by \citet{gil13a}.

At the very least, the real \boo\ stars have $[\alpha /Fe]=+0.3\pm 0.1$ dex, but this value seems to be unchanging; it is likely that SF terminated before the onset of SN Ia contamination. We assume this can be quantified as a colour-shift.
When we examined the 
Dartmouth isochrones, for $[\alpha/Fe]=0.0$ and $+0.4$, in the \str\ system only the y-filter seemed to be sensitive to $[\alpha/Fe]$, making stars brighter by 0.046 mag. at 12~Gyr and $[Fe/H]=-2.5$. We saw that the model stars with low metallicities lost sensitivity in the $v$- and $b$-filters, making
those magnitudes insensitive to $\alpha$-abundances. In the Washington filters, $T_1$ was shifted by -0.057 mag. and $(C-T_1)$ by +0.035, with a variation of less than 1 per cent . 

\begin{table*} 
\begin{minipage}{70mm}
\caption{Final Model Fits to \boo\ Stars}
\begin{tabular}{@{}lllll}
\hline 
{ID}&  {Age(Gyr)} &  {$[Fe/H]$}   &  {$T_{eff}(K)$} &  {$\log g$}\\
&	$\pm 0.5$ Gyr& $\pm 0.3$ dex& $\pm 50$ K& $\pm 0.10$ dex\\
\hline
\rowcolor{LC}
Boo-1137&   12.0&   -2.6&  4775&   1.34\\
\rowcolor{LC}
Boo-127&    10.8&   -2.2&  4743&   1.38\\
\rowcolor{LC}
Boo-117/HWB-8&      10.6&   -2.0&  4684&   1.42\\
\rowcolor{LC}
Boo-119/HWB-9&      10.8&   -1.8&  4705&   1.49\\
HWB-16&     11.5&   -1.6&  5022&   2.21\\
\rowcolor{Gray}
HWB-19&     11.5&   -3.4&  5292&   2.91\\
\rowcolor{LC}
HWB-22&     11.5&   -1.9&  5184&   2.60\\
\rowcolor{LC}
HWB-24&     11.5&   -2.4&  5314&   2.77\\
\rowcolor{LC}
HWB-28&     11.5&   -1.6&  5288&   2.89\\
HWB-29&     11.5&   -1.6& 5298& 2.92\\
HWB-31&     12.0&   -2.5&  5422&   3.03\\
\rowcolor{LC}
HWB-34&     11.5&   -1.7&  5365&   3.10\\
HWB-40&     11.6&   -1.6&  5411&   3.49\\
HWB-45&     12.0&   -3.0&  6919&   4.07\\
HWB-47&     11.9&   -1.7&  5574&   3.51\\
HWB-48&     11.5&   -3.5&  7022&   4.17\\
HWB-50&     11.5&   -1.8 & 5709&   3.62\\
HWB-51&     11.5&   -3.5&  6948&   3.91\\
\hline
\end{tabular}\\
$\chi^2$ fits to artificial stars generated from the $[\alpha/Fe]=0.0$ Dartmouth isochrones \citep{dot08}, using the $CT_1by$-filters.
Light cyan rows are confirmed members and the light gray row is a star that was ruled out by \cite{mar07} using its radial velocity.
\end{minipage}
\end{table*}

Setting $[\alpha/Fe]\sim +0.4$ for \boo , we compare our Class A--C stars'  $CT_1by$-magnitudes with the whole data set of over 50,000 artificial stars, and the results are listed in Table 9. 
We performed  $\chi^2$ fits to the  stars generated from the $[\alpha/Fe]=0.0$ Dartmouth isochrones \citep{dot08}, using the colour-shifts determined for the $CT_1by$-filters.
With the possible inclusion of two blue stragglers (HWB-45 and HWB-51), and HWB-9 being carbon-rich (and CN-strong), the average age is $11.5 \pm 0.4$~Gyr, and the range in metallicity is $-1.6>[Fe/H]>-3.4$, which would be shifted to around  $-1.9>[Fe/H]>-3.7$ with $\alpha$-enhancement added back \citep[in agreement with][]{gil13b}. We note that Boo-1137 does not appear as metal poor 
as it should, also, likely because the C-filter also loses sensitivity for very metal-poor stars. Our tests of the 
recovery of the artificial stars where we added typical photometric uncertainties found that the  $(C-T_1)$-colour started to become insensitive 
when $[Fe/H]<-2.2$. We also found it  necessary to use $(b-y)$ as the temperature-sensitive colour 
instead of $(T_1-T_2)$: otherwise, the temperatures of the real stars became at least 500K too hot compared to the SEDs shown in Figures 8--15. Using $m_{**}$ and $(b-y)$ or $(C-y)$ did not improve the fit. $(C-y)$ or $(C-T_1)$ alone are not sufficient for individual stars, unlike whole giant branches \citep{ros14}.

We conclude that the fewest number of filters we can use for these old, 
metal-poor populations, is 4, and those are $CT_1by$. We preserve the $\sim 2$ dex range in [Fe/H] from the spectra, but we estimate Table~9's $[Fe/H]$-values are, on average,  0.5 dex more metal-rich than \citet{lai11}.
Our results are more consistent with the CaT-calibrations, and the other spectroscopic studies \citep[see][]{gil12}.

To display the observational uncertainties in derived parameters for Table 9, Figure 17 shows the CMDs in \str , Washington, and SDSS filters for the model stars to match the real data shown in Figures 4 \& 18. Starting with the black circles for Case II, with $-4.0 < [Fe/H] <-1.0$ and ages, 10--14 Gyr. We then restrict the age range to 11-12 Gyr (green circles), and then to 11.3-11.7 Gyr (yellow). The green and yellow points have the same metallicity spread. Figure 17a--c are the models with no photometric scatter, and Figures 17d--f introduce the photometric uncertainty according to Figure 3 to match the \boo\ data (and if we had obtained similar SDSS-data instead of just the catalog magnitudes). We are able to detect the slight metallicity range contraction from the black points, $-4.0 < [Fe/H] <-1.0$, to the green points $-3.0 < [Fe/H] <-1.5$ at the SGB in Figure 17e, but not in the \str\ filters in Figure 17d, as expected from Figure 16 and Tables 7 \& 8. The vertical section of the MSTO is likewise able to detect the age contraction from a spread of 1~Gyr to 0.4~Gyr. We confirm that the SGB is sensitive to metallicity and not age, and the MSTO is more sensitive to age, except if we had very metal-poor with better than 1\%-photometry in $(u-g)$ at the MSTO.
We note that as the age spread contracts from the green to the yellow points for a few thousand stars in our models, and the RGB depopulates. The model-fits in Table 9 have a smaller value of $\chi^2$ when we use $(b-y)$ in addition to $(C-T_1)$, rather than $T_1-T_2$. We conclude that only combining $CT_1by$ can cover this range in $[Fe/H]$ and age smoothly, with no break between the RGB and the MSTO stars, unless it is caused by photometric uncertainties. The uncertainties in Table 9 metallicities are $\pm 0.2~dex$ in $[Fe/H]$ and $\pm 0.3$ Gyr in age using 4 filters and with 1\% photometry.

Figure 18 shows the final CMD with the $[\alpha/Fe]\sim +0.4$ Dartmouth isochrones. The Class A--C objects in \boo\  have an age
slightly less than 12~Gyr, but a wide [Fe/H]-range, likely exceeding 2 dex \citep{nor08}. It is
possible to use the $(C-T_1)$-colours alone, if the upper-RGB is well-populated and $[Fe/H]>-2.0$, to get age and metallicity \citep{ros14}. For most dSphs, with few
stars on the upper-RGB, the rRGB and MSTO stars are sufficient to break the age-metallicity degeneracy, if you have
2 colours, $(C-T_1)$) \& $(b-y)$.  Figure  17a shows that \str\ colours are not sufficient alone to extract age information from the
RGB with normal $\sim 1$ per cent  photometry, and that high S/N is necessary at the MSTO to recover age
information. We also note that Figure 18b shows  $(C-T_1)$ isochrones
for $-1.0>[Fe/H]>-2.5$ maintain good separation for ages 10--12~Gyr, only having a pinch-point at
the base of the RGB.

\section{Conclusions}

We have shown that the most efficient way to evaluate the metallicity, $[\alpha/Fe]$, and age of old, metal-poor systems 
is to observe individual stars with the $CT_1by$ filters combined with the Dartmouth isochrones. Application to the real \boo\ stars from Table 9 is as follows.
 If the distance and reddening are known, use the $T_1$-magnitude
and the $(C-T_1)$-colour to obtain $[Fe/H]$, $[\alpha/Fe]$, age, $\log T_{eff}$ and  $\log g$
from a $\chi^2$-fit to the Dartmouth $\alpha$-enhanced course model-grid, with ages of (10, 12,
\& 14~Gyr). Primarily, this
fixes $[\alpha/Fe]$ to $\pm 0.2$ dex for the star (or to see if $\alpha$-enhancement appears
constant for the system). Then we use another  $\chi^2$-fit  model stars' $T_1$-magnitudes,
$(C-T_1)$ \it and \rm $(b-y)$ colours, to
find the best-fit $[Fe/H]$,  age, $\log T_{eff}$ and  $\log g$ for each star. A recovery-test of this method 
returns over 90 per cent  of the model stars if we add in 1--5\% photometric errors.
Noting that spectroscopic surveys do not always agree with each other \citep{gil13a},
our final estimates of $[Fe/H]$ from photometry agree with 
the spectroscopy measurements within $\sim 0.2$ dex. If the model stars have $[Fe/H]=-1.0$ to $-2.0$,
then $T_1$-magnitudes and $(C-T_1)$-colours are sufficient, and the metallicity of the whole 
system can be defined by the $m_*$-index. If $[Fe/H]<-2.0$, the $(b-y)$-colours need to be added
and the metallicity alone can be found from $m_{**}$.

\begin{figure} 
\includegraphics[width=84mm,scale=1.0]{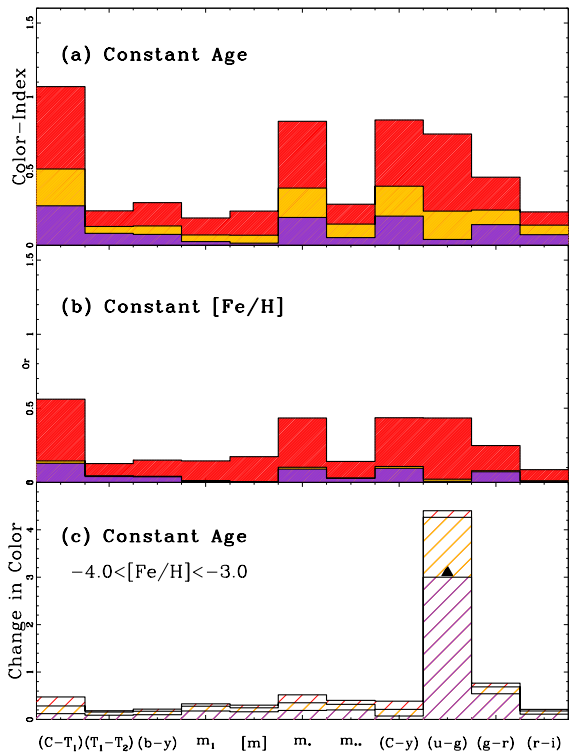}
\caption{Proportional bar chart for the colour \it ranges \rm for the models of RGB stars (red),  with $\log g$ of 1.0-2.0;  $\log g$ of 2.0-3.0 for the rising-RGB (rRGB; gold), below the level of the HB, and $\log g\sim 4.0$ for the MSTO (violet). The colours are shown for the \str , Washington and SDSS systems, at:
 \bf (a) \rm fixed age of 11.5~Gyr and $-2.5< [Fe/H]< -1.8$. 
Data from Table 7.
 \bf (b) \rm  Constant $[Fe/H]=-1.8$ (which would be 
equivalent to $\alpha$-enhanced more metal poor stars) and ages  10-14~Gyr.
 $(C-T_1)$ is most effective for all evolutionary stages in both age and metallicity, and rRGB colours are insensitive to age. Data from Table 8.
 \bf (c) \rm Colour-range for models with a fixed age of 11.5~Gyr and $-4.0< [Fe/H]< -3.0$. Note that $(u-g)\ge 3.0$ becomes extremely sensitive for extremely metal-poor MSTO stars.  }
 \end{figure}

\begin{figure} 
 \vskip 2cm
\includegraphics[width=84mm,scale=1.0]{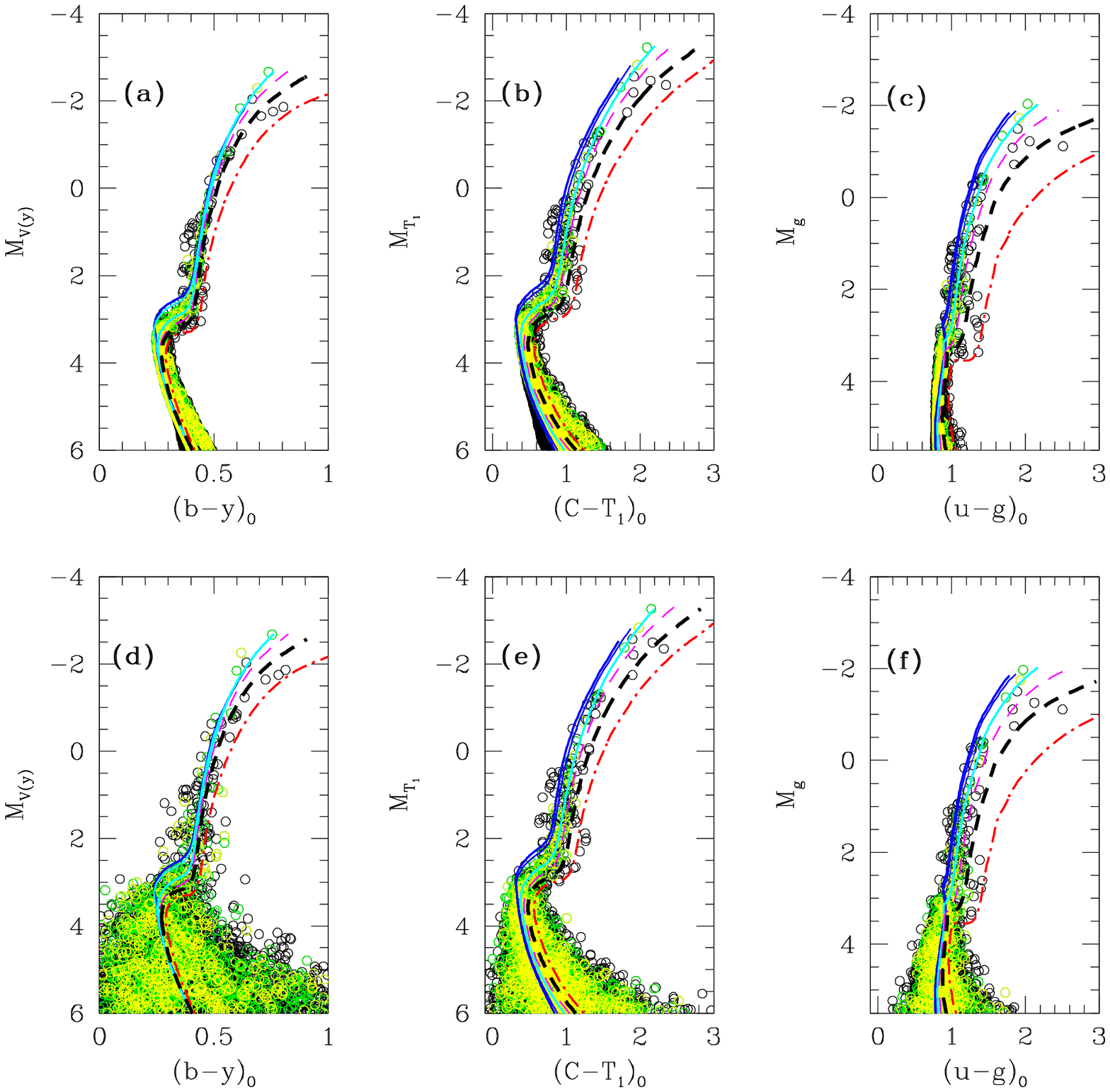}
\caption{We show
colour-magnitude diagrams for the \str  , Washington and SDSS data models. The Dartmouth isochrones for several values of $[Fe/H]$ and ages 10-12~Gyr are displayed. From left to right, the blue isochrones are
$[Fe/H]=-4.0$, -3.5, and -3.0, $[\alpha /Fe]=+0.0$, all with an age of 12~Gyr, decreasing in thickness. The cyan  line is $[Fe/H]=-2.5$, $[\alpha /Fe]=+0.4$ at 12~Gyr, the magenta dashed line is $[Fe/H]=-2.0$, $[\alpha /Fe]=+0.2$ at 12~Gyr, the black long-dashed line is $[Fe/H]=-1.5$, $[\alpha /Fe]=+0.0$, at 11~Gyr, and the red dot-dashed line is $[Fe/H]=-1.0$, $[\alpha /Fe]=+0.0$ at 10~Gyr. Here, the
black circles are the 9133 stars from Case II with with $-4.0 < [Fe/H] <-1.0$. The metallicity distribution  terminated after star formation lasting at least 4~Gyr. Green circles are 3187 stars with ages 11-12Gyr and $-3.0 < [Fe/H] <-1.5$. The yellow circles further restrict the age to 11.3-11.7 Gyr which shows the limits of age resolution for 1251 models stars.
isochrones for several values of $[Fe/H]$ are displayed
 \bf (a) \rm \str\ data for model stars with no added photometric scatter, but showing the broadened MS from the 0.5 binary fraction.
 \bf (b) \rm Washington CMD for the same sample.
 \bf (c) \rm  \str\ data with photometric scatter introduced according to Figure  3.
  \bf (d) \rm Washington CMD for the same sample, with the photometric scatter according to Figure 3.
 }
 \end{figure}

 \begin{figure} 
 \vskip 2cm
\includegraphics[width=84mm,scale=1.0]{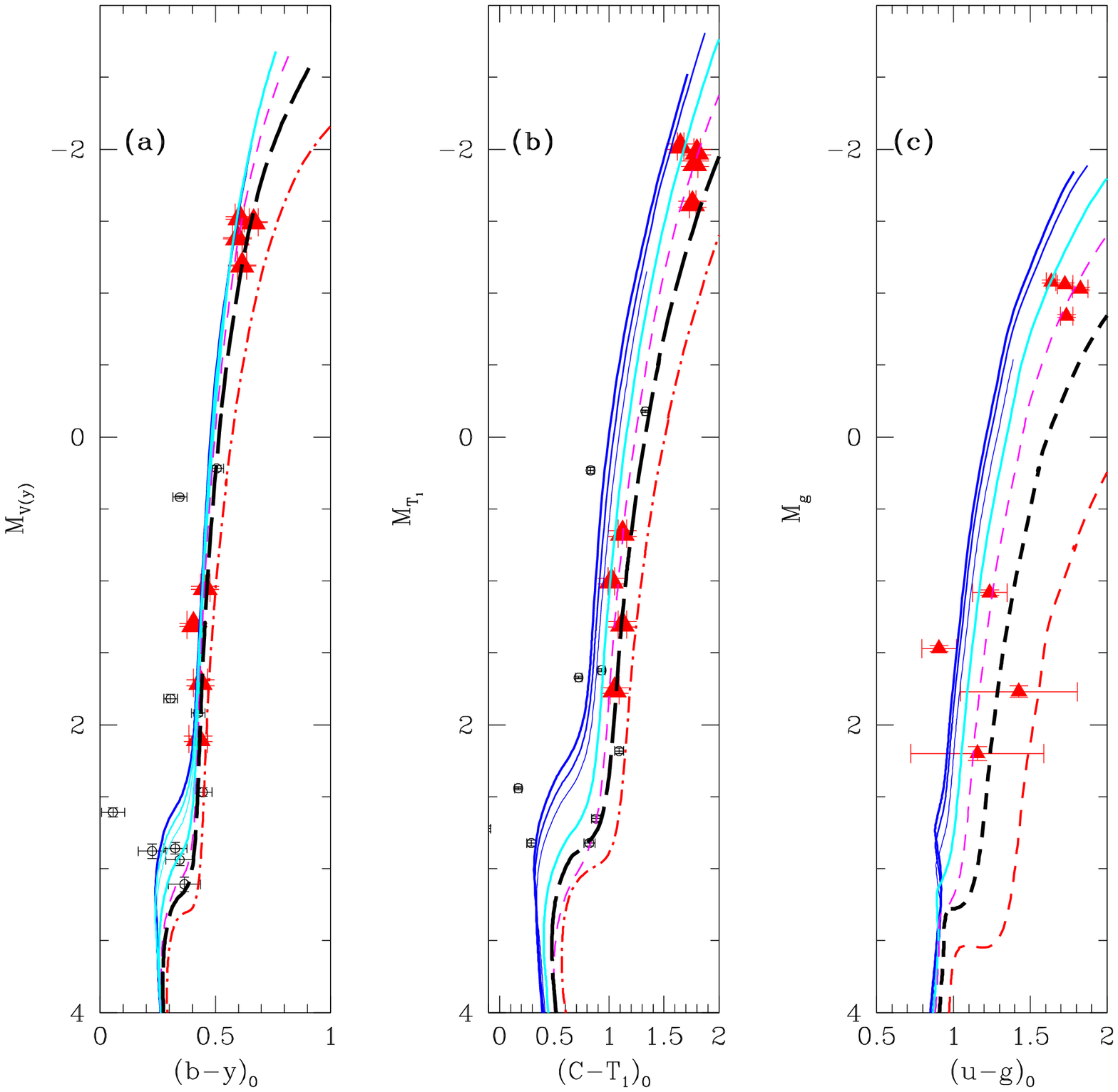}
\caption{We show colour-magnitude diagrams for the \str , Washington and SDSS data on \boo\ objects. The Dartmouth isochrones for several values of $[Fe/H]$ and ages 10-12~Gyr are displayed. From left to right, the blue isochrones are
$[Fe/H]=-4.0$, -3.5, and -3.0, $[\alpha /Fe]=+0.0$, all with an age of 12~Gyr, decreasing in thickness. The cyan  line is $[Fe/H]=-2.5$, $[\alpha /Fe]=+0.4$ at 12~Gyr, the magenta dashed line is $[Fe/H]=-2.0$, $[\alpha /Fe]=+0.2$ at 12~Gyr, the black long-dashed line is $[Fe/H]=-1.5$, $[\alpha /Fe]=+0.0$, at 11~Gyr, and the red dot-dashed line is $[Fe/H]=-1.0$, $[\alpha /Fe]=+0.0$ at 10~Gyr. In all plots, open circles are sources from Table~3 and the large red filled triangles represent the 8 proper motion-confirmed members. We use $E(B-V)=0.02$ and $DM=19.11$ for \boo .
 \bf (a) \rm The \str\ CMD with stars from Table~3. 
 \bf (b) \rm  The Washington CMD with stars from Table~3.
  \bf (c) \rm The SDSS CMD shown with 8 proper-motion members from Table 4.  }
 \end{figure}

Our \boo\ data in Figure 18 is thus consistent with the models given the photometric scatter and that we can resolve the  $[Fe/H]$-spread from at least -3.5 to -1.6, 
matching the ranges claimed by all the spectroscopic surveys. The large range in $[Fe/H]$, coupled with a small age range, which can be seen from combining the \str\ and Washington data, indicates a short SF history. 
The best-fit isochrones and Figure 18 give an age for \boo\ of $11.5\pm 0.4$~Gyr, confirming that the ISM was lost at very early times. The spectroscopic surveys are not spatially uniform, but we agree with \citet{gil13a,gil13b} and \citet{kop11} that \boo\ seems to be comprised of more than one population distribution. A change in the $\alpha$-element abundances is not seen in their data or our isochrone fits, so that star formation is predicted to last at least 0.5 Gyr before any SN Ia contribute. The CEMP-no object (Boo-119) was likely to have formed from the ejecta of Pop III stars \citep{gil13a,gil13b}, maybe only one \citetext{also see \citealp{kir12}}.

We recommend that Washington filters are used for the study of dSph systems beyond $\sim 50$ kpc (\boo\ is at about 65 kpc), except where  considerable reddening present; although they would still be better than $ugriz$. For future surveys using the Sloan filters, we recommend adding the Washington C-filter, as using  $(C-r)$ is the best (and cheapest) compromise for comprehensive stellar population studies, but that $(u-g)$ is very effective at the MSTO.  If a star
is not  carbon-enhanced, the $[Fe/H]$-sensitivity of the Washington and Str\"{o}mgren combination ($CT_1by$) is at least twice as great as that of the SDSS filters. Below the horizontal branch,
\str\ and Washington filter sets lose metallicity sensitivity, but $CT_1by$ succeeds where other calibrations fail. For upper-red-giant branch stars, the \str\ $m_1$-index gives a more-accurate metallicity estimate than the Washington filters compared to recent spectroscopic studies (around $\sim 0.3$ dex). Washington filters give better $[Fe/H]$-resolution for the range $-1.0>[Fe/H]>-2.0$, but for lower values, $[Fe/H]<-2.5$, $CT_1by$ is the most effective combination for these
populations.

\section*{Acknowledgments}
 This paper used observations obtained with the Apache Point Observatory 3.5-meter telescope, which is owned and operated by the Astrophysical Research Consortium. Hughes wishes to thank the APO observing staff for their support and late-night instant-messaging and Myra Stone for help with data analysis.
We also acknowledge financial support from the NSF, and the Kennilworth Fund of the New York Community Trust. We thank Frank Grundahl (for sharing his M92 data), and Myra Stone for help  with DAOPHOT III. We are grateful to: Ata Sarajedini, Inese Ivans, Peter Stetson, 
Jeff Brown,  Beth Willman, John Norris, Anna Frebel, Marla Geha, Ricardo Munoz, Kim Venn, and Ryan Leaman for useful  discussions. We made use of the SDSS DR7:
Funding for the Sloan Digital Sky Survey (SDSS) and SDSS-II has been provided by the Alfred P. Sloan Foundation, the Participating Institutions, the National Science Foundation, the U.S. Department of Energy, the National Aeronautics and Space Administration, the Japanese Monbukagakusho, and the Max Planck Society, and the Higher Education Funding Council for England. The SDSS Web site is http://www.sdss.org/.  The SDSS is managed by the Astrophysical Research Consortium (ARC) for the Participating Institutions. The Participating Institutions are the American Museum of Natural History, Astrophysical Institute Potsdam, University of Basel, University of Cambridge, Case Western Reserve University, The University of Chicago, Drexel University, Fermilab, the Institute for Advanced Study, the Japan Participation Group, The Johns Hopkins University, the Joint Institute for Nuclear Astrophysics, the Kavli Institute for Particle Astrophysics and Cosmology, the Korean Scientist Group, the Chinese Academy of Sciences (LAMOST), Los Alamos National Laboratory, the Max-Planck-Institute for Astronomy (MPIA), the Max-Planck-Institute for Astrophysics (MPA), New Mexico State University, Ohio State University, University of Pittsburgh, University of Portsmouth, Princeton University, the United States Naval Observatory, and the University of Washington.

\nocite{mun06}
\nocite{ven04}
\nocite{ven08}
\newpage
\bibliography{boowork5}

\end{document}